\newcommand{\paperversion}{arxiv}
\def\anonymoussubmission{1} \RequirePackage{paperversions}
 \documentclass[conference,compsoc]{IEEEtran}

\RequirePackage[numbers]{natbib}
\RequirePackage{amsthm}
\RequirePackage{capt-of}
\RequirePackage{amssymb}
\newtheorem{theorem}{Theorem}[section]
\newtheorem{lemma}[theorem]{Lemma}
\theoremstyle{definition}
\newtheorem{definition}[theorem]{Definition}

\makeatletter
\renewcommand\paragraph{\@startsection{paragraph}{4}{\z@}{0.5\baselineskip plus 0.25\baselineskip minus 0.2\baselineskip}{-1em}{\normalfont\normalsize\bfseries}}
\makeatother

\newcommand{\paperfrontmatter}{\maketitle

  \begin{abstract}
    The real--ideal paradigm, an approach to security specification
developed in the cryptography community, can express sophisticated
confidentiality and integrity requirements.
We show that this paradigm is the basis for a simple yet powerful unifying semantic
framework for computer systems security.
Our primary case study is the family
of security properties related to nonmalleable information flow (NMIF).
Although NMIF has proved useful in building secure decentralized systems,
these properties---which involve information flow downgrading---are not well
understood because past semantics based on hyperproperties are complex
and intensional---tied to the type systems that enforce them.
Using the real--ideal paradigm, this work clarifies the role of prior type
systems as well as the semantic properties they enforce: (1)~semantic security
properties are purely extensional real--ideal statements; (2)~type system
soundness too, surprisingly, is another real--ideal statement, which implies
semantic security for the entire NMIF family. Although clarifying NMIF is the
primary focus, we also discuss how other frameworks and properties such as
trace refinement, relaxed noninterference, and delimited release may be
captured in the real--ideal paradigm. Additionally, by incorporating expressive
confidentiality policies, our model goes beyond previous applications of the
real--ideal paradigm outside cryptography (such as to hardware and distributed
systems). All results, including an implementation of a mechanically verified
IFC type checker, are fully formalized using the Rocq prover.
   \end{abstract}

  \iffinalformat
      Keywords:
  \fi
}

\newcommand{\paperappendix}{\appendices
  \crefalias{section}{appendix}\crefalias{subsection}{subappendix}}

\RequirePackage{graphicx, color, eso-pic}
\RequirePackage{amsmath, stmaryrd, utf8math, checkpagelimit}
\RequirePackage[T1]{fontenc}
\RequirePackage{inconsolata}
\pagelimit{13}
\ifshowcomments
  \usepackage{authcomments}
\else
  \usepackage[disable]{authcomments}
\fi
\RequirePackage{hyperref}
\hypersetup{
    colorlinks,
    urlcolor=black,
    citecolor=black,
    filecolor=black,
    linkcolor=black
}
\RequirePackage{cleveref}
\RequirePackage{etoolbox}

\AtBeginEnvironment{definition}{\crefalias{theorem}{definition}}
\crefname{definition}{definition}{definitions}
\Crefname{definition}{Definition}{Definitions}
\AtBeginEnvironment{lemma}{\crefalias{theorem}{lemma}}
\crefname{lemma}{lemma}{lemmas}
\Crefname{lemma}{Lemma}{Lemmas}

\RequirePackage{aplcomments}
\RequirePackage{pl-syntax}
\RequirePackage{listings}
\lstdefinestyle{ocamlish}{
  basicstyle=\ttfamily\small,
  keywordstyle=\bfseries,
  keywords={def,match,with,end,return,endorse},
  morecomment=[l]{//},
  mathescape=true,
  columns=fullflexible,
}
\RequirePackage{fontawesome5}
\RequirePackage{tikz}
\usetikzlibrary{arrows.meta, calc, patterns, fit, positioning}
\RequirePackage{tikz-cd}
\RequirePackage{multirow}
\RequirePackage{judgments}
\RequirePackage{functions}
\RequirePackage{inference-rules}

\ifshowcomments
 \ifwarning
  \AtBeginDocument{
    \AddToShipoutPicture*{\put(506,720)
      {\it\color{red}{\fbox{\large Draft---please do not distribute}}}}
  }
 \fi
\fi

\ifblinded
  \author{}
\else
  \author{\IEEEauthorblockN{Suraaj Kanniwadi and Andrew C. Myers}
    \IEEEauthorblockA{Cornell University\\ \{suraaj,andru\}@cs.cornell.edu}}
\fi
  \RequirePackage{xspace}

\newcommand{\emulates}{\xrightarrow{\text{emulates}}}
\newcommand{\sembr}[1]{\llbracket #1 \rrbracket}
\newcommand{\defined}{\triangleq}
\newcommand{\impthreat}{\textsc{ImpThreat}\xspace}

\newcommand{\setfnt}[1]{\ensuremath{\mathrm{#1}}\xspace}
\newcommand{\mvarfnt}[1]{\ensuremath{#1}\xspace}
\newcommand{\kwfnt}[1]{\ensuremath{\mathtt{#1}}\xspace}
\newcommand{\fnfnt}[1]{\ensuremath{\mathit{#1}}\xspace}

\newcommand{\circnum}[1]{\tikz[baseline=(n.base)]{\node[circle, fill=black, text=white,
    inner sep=1pt, minimum size=1em, font=\scriptsize\sffamily] (n) {#1};}}

\newcommand{\advdevil}[1][0.5]{\ensuremath{\text{\raisebox{0.15ex}{\scalebox{#1}{\faSkullCrossbones}}}}\xspace}

\newcommand{\stparr}{\xrightarrow{\setfnt{store}}}
\newcommand{\cntarr}{\xrightarrow{\setfnt{cont}}}
\newcommand{\evarr}{\xrightarrow{\setfnt{event}}}

\newcommand \squigeq {\approx}
\newcommand \squigeqt {\approx^{t}}
\newcommand \ssquigeq {\approx_{s}} 
\newcommand \wsquigeq {\approx_{w}} 
\newcommand \ssquigeqt {\approx^{t}_{s}} 
\newcommand \wsquigeqt {\approx^{t}_{w}}

\newcommand \prog {\mvarfnt {p}} \newcommand \realsys {\mvarfnt {s_r}}
\newcommand \atk {\mvarfnt {A}}
\newcommand \realatk {\mvarfnt {A_r}}
\newcommand \idealatk {\mvarfnt {A_i}}
\newcommand \simatk {\mvarfnt {A_{sim}}}
\newcommand \idealsys {\mvarfnt {s_i}}
\newcommand \aliceatkset {\mvarfnt {Atk_{alice}}}
\newcommand \bobatkset {\mvarfnt {Atk_{bob}}}
\newcommand \ratkset {\setfnt{Atk_{real}}} 
\newcommand \iatkset {\setfnt{Atk_{ideal}}} 
\newcommand \dgatkset {\setfnt{Atk_{dgrst}}}
\newcommand \nmifatkset {\setfnt{Atk_{nmif}}}
\newcommand \rdatkset {\setfnt{Atk_{rd}}}
\newcommand \teatkset {\setfnt{Atk_{te}}}
\newcommand \niatkset {\setfnt{Atk_{ni}}}

\newcommand{\extpair}{\fnfnt{extract_{pair}}}
\newcommand{\extreal}{\fnfnt{extract_{real}}}
\newcommand{\extideal}{\fnfnt{extract_{ideal}}}
\newcommand{\stepcomb}{\fnfnt{step_{comb}}}
\newcommand{\steppair}{\fnfnt{step_{pair}}}
\newcommand{\stepreal}{\fnfnt{step_{real}}}
\newcommand{\stepideal}{\fnfnt{step_{ideal}}}
\newcommand{\filterideal}{\fnfnt{filter_{ideal}}}
\newcommand{\idfn}{\fnfnt{id}}

\newcommand{\tracea}{\ensuremath{\tau_1}\xspace}
\newcommand{\traceb}{\ensuremath{\tau_2}\xspace}
\newcommand{\vleak}{\mvarfnt{v_{\mathit{leak}}}}

\newcommand{\stadvset}{\setfnt{State_{adv}}}
\newcommand{\initadv}{\mvarfnt{adv_{init}}}
\newcommand{\stepadv}{\mvarfnt{adv_{step}}}
\newcommand{\cmdpre}{\mvarfnt{cmd_{pre}}}
\newcommand{\cmdpost}{\mvarfnt{cmd_{post}}}

\newcommand{\ba}{\mvarfnt{b_1}}
\newcommand{\bb}{\mvarfnt{b_2}}
\newcommand{\initcfg}{\mvarfnt{\kappa_{init}}}
\newcommand{\siginp}{\mvarfnt{\sigma_{inp}}}
\newcommand{\evc}{\mvarfnt{ev_c}}
\newcommand{\eva}{\mvarfnt{ev_a}}
\newcommand{\termev}{\kwfnt{\downarrow}}

\newcommand{\cmdendorse}{\kwfnt{endorse}} \newcommand{\cmddeclassify}{\kwfnt{declassify}}

\newcommand{\judgec}{\ensuremath{\mvarfnt{pc} \wts \mvarfnt{c}}\xspace}
\newcommand{\judgep}{\ensuremath{\mvarfnt{pc} \wts \prog}\xspace}
\newcommand{\wts}{\vdash_{\advpower[0.5]}}   \newcommand{\advpower}[1][0.7]{\ensuremath{\text{\scalebox{#1}{\textcolor{black!70}{\faSkull}}}}\xspace}
\newcommand{\advpowerof}[1]{\advpower(#1)}
\newcommand{\lbl}{\ensuremath{\ell}\xspace}
\newcommand{\lblof}[1]{\ensuremath{\lbl_{#1}}\xspace}
\newcommand{\srcfn}{\setfnt{src}}
\newcommand{\dstfn}{\setfnt{dst}}
\newcommand{\PT}{\ensuremath{\mathtt{PT}}\xspace}
\newcommand{\PU}{\ensuremath{\mathtt{PU}}\xspace}
\newcommand{\ST}{\ensuremath{\mathtt{ST}}\xspace}
\newcommand{\SU}{\ensuremath{\mathtt{SU}}\xspace}
\newcommand{\flowto}{\ensuremath{\sqsubseteq}\xspace}
\newcommand{\lbljoin}{\ensuremath{\sqcup}}
\newcommand{\lblmeet}{\ensuremath{\sqcap}}
\newcommand{\wt}{\vdash}
\newcommand{\conf}[1]{#1^{\rightarrow}}
\newcommand{\integ}[1]{#1^{\leftarrow}}
\newcommand{\uncomp}[1]{\blacktriangledown{#1}}
\newcommand{\stronger}{\ensuremath{\succcurlyeq}}

\newcommand{\syscont}{\mvarfnt{k}}
\newcommand{\syscontrl}{\mvarfnt{k_{real}}}
\newcommand{\syscontidl}{\mvarfnt{k_{ideal}}}

\newcommand{\sigsim}{\mvarfnt{\sigma_{sim}}}
\newcommand{\sigidl}{\mvarfnt{\sigma_{idl}}}

\newcommand{\pisim}{\mvarfnt{\pi_{sim}}}
\newcommand{\advst}{\mvarfnt{ast}}
\newcommand{\cmd}{\mvarfnt{c}}
\newcommand{\cmdr}{\mvarfnt{c_{r}}}

\newcommand{\configuration}[3]{\ensuremath{\langle#1, #2, #3\rangle}}

\newcommand{\storeset}{\setfnt{Store}}
\newcommand{\natset}{\setfnt{Nat}}

\newcommand{\pcset}{\setfnt{PC}}
\newcommand{\cmdcmbset}{\setfnt{Cmd_{cmb}}}
\newcommand{\contcmbset}{\setfnt{Cont_{cmb}}}
\newcommand{\stcmbset}{\setfnt{State_{cmb}}}
\newcommand{\pc}{\mvarfnt{pc}}
\newcommand{\cmdcmb}{\mvarfnt{cmd_{cmb}}}
\newcommand{\cropt}{\mvarfnt{c^?_r}}
\newcommand{\ciopt}{\mvarfnt{c^?_i}}
\newcommand{\kcmb}{\mvarfnt{k_{cmb}}}
\newcommand{\stcmb}{\mvarfnt{st_{cmb}}}

\newcommand{\sigr}{\mvarfnt{\sigma_r}}
\newcommand{\sigi}{\mvarfnt{\sigma_i}}
\newcommand{\sigs}{\mvarfnt{\sigma_s}}

\newcommand{\trusted}{\fnfnt{trusted}}
\newcommand{\public}{\fnfnt{public}}

\newcommand{\proj}[2]{#1|_{#2}}

\newcommand {\invcomb} {\fnfnt{Inv_{comb}}}

\newcommand{\configcmb}{\mvarfnt{\kappa_{cmb}}}
\newcommand{\configpair}{\mvarfnt{\kappa_{pair}}}
\newcommand{\configreal}{\mvarfnt{\kappa_{real}}}
\newcommand{\configideal}{\mvarfnt{\kappa_{ideal}}}

\newcommand{\cfgcmbset}{\setfnt{Cfg_{cmb}}}
\newcommand{\cfgpairset}{\setfnt{Cfg_{pair}}}
\newcommand{\cfgrealset}{\setfnt{Cfg_{real}}}
\newcommand{\cfgidealset}{\setfnt{Cfg_{ideal}}}

\newcommand{\Pset}{\mvarfnt{P}}

 \title{Demystifying Information Flow Downgrading using~the~Real--Ideal~Paradigm}
 
\begin{document}
\paperfrontmatter
\section{Introduction}
\label{sec:Intro}

Building secure systems is difficult: they must provide some degree of
functional correctness even in the presence of adversaries who try
to subvert their functioning or learn unauthorized secrets. A
serious obstacle to designing secure systems is that we lack
security specifications: precise descriptions of what it means for the
system to be secure. Without such specifications, arguments about
whether the system is secure have no solid foundation, and we have no
way to prove that systems are secure. In the current era where
automated AI-driven attacks pose a growing threat, security proofs
may be the best hope for trustworthy computing systems.

There have been many attempts to build systems that offer strong
security guarantees by construction. In particular, information flow
properties are increasingly recognized as crucial, and systems like
DStar~\cite{dstar}, Fabric~\cite{fabric09}, and LIO~\cite{lio} aim to
control insecure information flow. However, the flexibility to build
practical applications requires downgrading mechanisms to allow the
controlled release of sensitive information and the controlled
acceptance of untrustworthy information.  What is lacking is an
adequate way to specify the security these systems deliver in the
presence of downgrading.

An important property for any specification is to be
\emph{extensional}, meaning that the specification should constrain the
externally observable behavior of the system rather than how the
system is expressed as code. A standard example of an extensional
security property is noninterference~\cite{GM82}, the \emph{E.  coli}
of information-flow security properties---highly studied but too
simple. However, noninterference does not capture the nuances of real
computer systems, because it disallows downgrading entirely; more
expressive power is needed to capture the subtle security
requirements of real applications.

Therefore, this paper advocates for a new framework for extensionally
specifying language-based security, adapted from the cryptographic
\emph{real--ideal paradigm}~\cite{gmw87, UC, constructivecrypto,
Maurer2011AbstractC}.\SK{Several? Our main case study is NMIF... }
We demonstrate how several existing
information-flow properties are given simple, intuitive descriptions
in this framework. Noninterference is neatly captured by this
framework, but so are less-understood properties such as robust
declassification~\cite{zm01b} (RD) and nonmalleable information flow
(NMIF)~\cite{nmifc}.  The new formal description captures NMIF more
simply and extensionally, while yielding new insights into why NMIF is
useful in practice.  This new approach improves on the previous
specification of NMIF as a 4-hyperproperty, because that specification
is \emph{intensional}, requiring both a semantics instrumented with
syntactic events and a complex exclusion of \emph{inept attackers}.

Specifying security is valuable, but we also need to be able to
soundly \emph{enforce} security properties.  The real--ideal paradigm is
useful here as well: even the soundness statements for practical
information flow type systems can be given as real--ideal statements,
and the technique is flexible enough to handle \emph{downgrading},
which practical information flow type systems must support.
Information flow control (IFC) advocates have argued that explicit
downgrading mechanisms allow programmers to focus their security
reasoning on the uses of downgrading, which can then be manually
audited to detect security vulnerabilities.  This argument has
significant empirical support~\cite{ifappstore}, but solid theoretical
underpinnings have been lacking.  We formalize this argument as a
novel real--ideal statement: an IFC type system enforcing NMIF ensures that
that a real attacker is no more powerful than an weak attacker who can
only interact with the system at downgrades.

To formalize this new security framework, we define a simple imperative
language \impthreat whose operational semantics are enriched with adversary
interactions. Our real--ideal emulation definition for \impthreat allows
simple, extensional definitions of information flow security properties.
Real--ideal IFC type soundness is proved for an \impthreat information flow
type system. In fact, this type system also enforces \emph{heterogeneous
security}: a generalization that ensures security for systems implemented by
mutually untrusting principals.

Our contributions are summarized as follows:motivating a need for better security specifications (\cref{sec:Motivation,sec:Information flow security}), 
a language-based formulation of the real--ideal paradigm (\cref{sec:The real--ideal paradigm,sec:Semantic framework}),
a real--ideal definition of sound type-based IFC enforcement (\cref{sec:Real--ideal soundness for information flow types}),
a formally verified type system satisfying real--ideal soundness (\cref{sec:Simplified NMIFC,sec:heterogeneous NMIFC}), and
a discussion of other properties which can be expressed in the real--ideal paradigm (\cref{sec:Beyond NMIF}).

All formalization is fully mechanized and verified in the Rocq theorem prover.
No admits or non-standard axioms are used.
The formalization is available at 
\iffinalformat \url{https://github.com/kssuraaj28/demystify-ifc-rocq} \relax
\else \url{https://anonymous.4open.science/r/demystify-ifc-rocq-8CB2/} \fi.

\section{Motivation}
\label{sec:Motivation}

\subsection{Specifying and enforcing security requirements}
\label{sub:Specifying and Enforcing Security Requirements}

Trace properties~\cite{liveness,tla} are the standard way to specify
allowed system behavior. A trace property is a collection of permitted
single executions; an implementation satisfies a specification if each
individual execution of the implementation is permitted by the specification. In
other words, the implementation behaviors \emph{refine} the behaviors allowed
by the specification. Such specifications are very useful when reasoning about
functional correctness but are not expressive enough to capture security
requirements.

This lack of expressiveness is because many natural security properties
(especially those involving privacy) talk about causality or information flow,
which cannot be expressed as trace properties~\cite{Schneider-EM}.
Admittedly, the notion of
\emph{hyperproperties}~\cite{cs08} addresses this limitation of trace
refinement, and generalizes the notion of what specifications are and
what it means for an implementation to satisfy a specification by
defining a specification to simply be a set of implementations (trace sets) that
satisfy it. Hyperproperties are a general description of allowed
behavior---too general, perhaps, because they encompass behavioral restrictions that
we do not normally think of as being about security. There is a burgeoning
literature on techniques for logically specifying and model checking
hyperproperties~\cite{cfkmr14,dm24,frs15},
but there are real gaps both between translating human-understandable
security requirements into these logics and between these logics and
in their practical enforcement.

The cryptography community offers another solution to the
problem of stating and enforcing security specifications:
the real--ideal paradigm. 
In this paradigm, systems are enriched with an explicit, operational
\emph{threat model}: back-door interfaces via which adversaries can
learn information and influence the system. A \emph{specification} is
also modeled as such a system---the \emph{ideal system}---which
is simpler and has more-constrained adversary interfaces. The
goal is then to demonstrate that this ideal system is \emph{emulated}
by the real system, by showing that every attack on
the real system is equivalent to some attack on the
ideal system.

While the details are made precise in \cref{sec:The real--ideal paradigm},
it is important to understand that this equivalence includes not only
regular system input--output (I/O) but also \emph{adversary leakage}:
information that the adversary learns. System I/O equivalence captures integrity
requirements, and adversary leakage equivalence captures confidentiality
requirements. Crucially, real--ideal emulation can capture realistic system
specifications which intertwine integrity and confidentiality requirements, such
as \emph{nonmalleable information flow} (NMIF)~\cite{nmifc} and \emph{robust
declassification}~(RD)~\cite{zm01b}.
\SK{Andrew says "reads like prior work"}

\subsection{NMIF as real--ideal emulation}
\label{sub:Making sense of nonmalleable information flow}
\begin{figure*}
  \centering
  \begingroup

\newcommand{\statesep}{1cm}   \newcommand{\lblxoff}{0.35}   \newcommand{\lblinner}{0.6}   \newcommand{\lblstep}{0.35}   \newcommand{\panelgap}{0.8em} 

\tikzset{
  cell/.style={fill=black!10, rounded corners=2pt, inner sep=2pt,
               text height=1.2ex, text depth=0.4ex, font=\large},
  outerstate/.style={draw, rounded corners=3pt, inner sep=3pt},
  arr/.style={-{Stealth[length=2.5mm]}, thick},
  arrnum/.style={midway, above, font=\scriptsize},
  ptrlbl/.style={font=\scriptsize, anchor=east},
  ptrline/.style={thin, dashed},
}
\newcommand{\shadedcell}[1]{\tikz[baseline=(c.base)]{\node[cell] (c) {\ensuremath{#1}};}}
\newcommand{\statequad}[4]{\tikz[baseline=(o.base)]{\node[outerstate] (o) {\shadedcell{#1}\,\shadedcell{#2}\,\shadedcell{#3}\,\shadedcell{#4}};}}

\noindent
\begin{minipage}[t]{.45\linewidth}
\vspace{0pt}
\begin{lstlisting}
def auctioneer (alice_bid, bob_bid):
  a := endorse (alice_bid)
  b := endorse (bob_bid)
  return match (compare a b) with
         | Gt => "Alice wins"
         | Lt => "Bob wins"
         end
\end{lstlisting}
\end{minipage}\hspace{\panelgap}\begin{minipage}[t]{.38\linewidth}
\vspace{0pt}
\textbf{Labels:} $A$ (Alice), $B$ (Bob), $A \wedge B$ (Both)

\bigskip

\centering
\begin{tabular}{|c|c|c|c|}
\hline
\multirow{2}{*}{Variable} & \multirow{2}{*}{Label\textsubscript{int}} & \multicolumn{2}{c|}{Label\textsubscript{conf}} \\
\cline{3-4}
                       &                                 & insecure           & secure                                \\
\hline
\texttt{alice\_bid}    &   $A$                           &  $A \wedge B$      &  $A$                                  \\
\texttt{bob\_bid}      &   $B$                           &  $A \wedge B$      &  $B$                                  \\
\hline
\end{tabular}

\end{minipage}

\bigskip

\begin{tikzpicture}[font=\small]

\node[cell] (c1a) at (0,0)           {$v_a$};
\node[cell, right=1.5pt of c1a] (c1b) {$v_b$};
\node[cell, right=1.5pt of c1b] (c1c) {$0$};
\node[cell, right=1.5pt of c1c] (c1d) {$0$};
\node[outerstate, fit=(c1a)(c1d)] (s1) {};

\coordinate (lblx) at ($(s1.west)+(\lblxoff,0)$);
\node[ptrlbl] (lbb) at ($(lblx)+(0, { \lblinner+\lblstep})$)  {\lstinline|bob_bid|};
\node[ptrlbl] (lba) at ($(lblx)+(0, { \lblinner})$)           {\lstinline|alice_bid|};
\node[ptrlbl] (la)  at ($(lblx)+(0, {-\lblinner})$)           {\lstinline|a|};
\node[ptrlbl] (lb)  at ($(lblx)+(0, {-\lblinner-\lblstep})$)  {\lstinline|b|};

\draw[ptrline] (c1a.north) |- (lba.east);
\draw[ptrline] (c1b.north) |- (lbb.east);
\draw[ptrline] (c1c.south) |- (la.east);
\draw[ptrline] (c1d.south) |- (lb.east);

\node[right=\statesep of s1] (s2) {\statequad{v_a}{v_a{+}1}{0}{0}};
\draw[arr] (s1.east) -- (s2.west) node[arrnum] {\protect\circnum{1}};

\node[right=\statesep of s2] (s3) {\statequad{v_a}{v_a{+}1}{v_a}{v_a{+}1}};
\draw[arr] (s2.east) -- (s3.west) node[arrnum] {\protect\circnum{2}};

    \node[below=0pt of s3] (actions) {
      \begin{tabular}{@{}l@{}}
        \circnum{1}\quad \advdevil[0.7]-Bob sets \lstinline|bob_bid := alice_bid + 1|. \\
        \circnum{2}\quad Bids endorsed by \lstinline|auctioneer|.
      \end{tabular}
    };
\end{tikzpicture}
\endgroup
   \caption{An insecure sealed-bid auction (top) and its attack trace under Bob's mauling (bottom).}
  \label{fig:auction}
\end{figure*}

Nonmalleable information flow (NMIF) was originally
motivated~\cite{nmifc} using a sealed-bid auction
(\cref{fig:auction}).
Here, Alice and Bob engage in a two-party sealed-bid auction implemented by
\lstinline |auctioneer|. They both trust \lstinline|auctioneer| to take their
bids independently and to reveal the winner. The labels $A$ and $B$ denote the
trust levels of Alice and Bob respectively, and thus the input bids that they
supply, \lstinline|alice_bid| and \lstinline|bob_bid|, have those integrity
labels. Since \lstinline |auctioneer| is written using a language that enforces
information flow control, it must downgrade these untrusted input bids using
\lstinline|endorse| expressions.

An \lstinline|auctioneer| that violates NMIF allows an attack where a malicious
Bob sets his bid based on Alice's: \lstinline|bob_bid := alice_bid + 1|, thereby
winning the auction with a minimal winning bid. This insecure program has
the confidentiality labels of the input bids set to $A \wedge B$---a highly
restrictive confidentiality label that neither Alice nor Bob can read---which
permits the assignment \lstinline|bob_bid := alice_bid + 1| without violating
the rules for secure information flow. To satisfy NMIF, \lstinline|auctioneer|
lowers the confidentiality label for the input bids as shown in the
figure, and \lstinline|bob_bid := alice_bid + 1| is no longer
permitted because $A$ confidentiality data (\lstinline|alice_bid|) is 
assigned into a $B$ confidentiality variable (\lstinline|bob_bid|).
The higher confidentiality label for the bids makes the program less secure!

The original paper pins this program's violation of NMIF to insecure
\lstinline|endorse| statements, where the source variables of the endorsement
contain confidential information \emph{inaccessible} to the provider. The
insecure \lstinline|auctioneer| endorses \lstinline|bob_bid| containing
information about \lstinline|alice_bid|, which is inaccessible to Bob. NMIF
is formalized using \emph{endorsement events} and \emph{declassification
events} that explicitly appear in the program trace. Because it depends on these
program events, the original NMIF formalization is not fully \emph{extensional}.
Extensional definitions---like noninterference---are preferable because
they abstract away from implementation details and depend only on 
observable system behavior.

We can formulate NMIF extensionally. Our insight is that the
program's insecurity arises from the advantage an adversary gains by
\emph{mauling}---changing untrusted state as a function of other (possibly
secret) system state. Bob updates the variable \lstinline|bob_bid| using state
(\lstinline|alice_bid|) that he cannot see, letting him win the auction with a
minimal winning bid.
\SK{Andrew says: general connection? not just to this program}

A secure program satisfying NMIF should not give such maulers any advantage. We
can state this as an extensional security requirement: programs satisfy NMIF if
an adversary that can maul is equivalent in power to one that cannot. This is a
real--ideal statement where ideal adversaries cannot maul and real adversaries
may.
\SK{Forward reference to more detail?}

\subsection{Information flow type soundness as real--ideal emulation}
\label{sub:Real--ideal information flow type soundness}

Practical information flow type systems~\cite{myers-popl99, ps03, Paragon}
allow downgrading. Unfortunately, with downgrading, noninterference is lost.
Despite losing this formal guarantee, in practice, such type systems seem to
provide some informal security assurance to programmers, allowing them to focus
on these downgrading statements when looking for program vulnerabilities. Empirical
studies of using static information-flow checks along with downgrading suggest
that this assurance is real~\cite{ifappstore}. It seems to suffice to reason
about adversarial interaction with only program downgrading statements, even
though real-world adversaries seem to have more power.

Despite the empirical evidence supporting this folk theorem, there is no formal
statement backing it. Such a statement is crucial to have a building block for
provably enforcing security of systems with downgrading. For instance, as we
show in \cref{sub:Implications}, such a  security requirement allows us to
prove program properties such as NMIF.

The real--ideal paradigm again naturally captures the requirement that it
suffices to reason about downgrade-constrained adversaries: ideal adversaries
are downgrade-constrained, while real adversaries are not. We formally specify
this requirement of downgrade-constrained adversaries in \cref{sec:Real--ideal
soundness for information flow types}. Additionally, we demonstrate a
mechanically verified enforcement (\cref{sec:Simplified
NMIFC,sec:heterogeneous NMIFC}) using a static information flow type system
similar to NMIFC~\cite{nmifc} that permits controlled downgrading.
Well-typed programs satisfy a real--ideal soundness theorem: powerful real-world
adversaries are simulated by downgrade-constrained ideal-world adversaries.
This type system is also designed with the goal of \emph{heterogeneous security}, where
mutual distrust between principals requires us to consider and enforce
security for multiple adversarial scenarios.

The Viaduct system~\cite{viaduct-pldi21, viaduct-formal} builds on a
similar insight about the value of real--ideal emulation in the setting of
language-based security, but for a limited language that does not permit
secret or untrusted control flow. The Viaduct proof of correctness is phrased
as real--ideal emulation connecting its source language to a distributed,
concurrent, cryptographic target language. The proof relies on the fact that the
Viaduct source language subjects downgrading to a similar typing discipline,
but the role of this restriction is not made clear. The present work has a more
general source language, and its real--ideal emulation clarifies the role of
information-flow typing in Viaduct (\cref{sub:Connection to Viaduct}).
\SK{Andrew says story not clear}

\section{The real--ideal paradigm}
\label{sec:The real--ideal paradigm}

\SK{Andrew says something is unclear here. I wonder what}
Informally, the real--ideal paradigm can be thought of as a generalization of
\emph{trace refinement}~\cite{abadi92,lynch-tuttle}. While the goal of
refinement is to show how a system implementation \emph{refines} a more
abstract specification, the real--ideal paradigm requires an \emph{emulation}
from a more abstract specification \emph{along with a more abstract threat
model}. In the abstract threat model, the adversary lacks some power that the
real adversary possesses; the emulation shows that the real adversary gains no
advantage from possessing that power.

\begin{definition}[Real--ideal emulation]
\label{def:risim}
A real system \realsys \emph{emulates} an ideal system \idealsys (denoted $
\realsys \emulates \idealsys $) if every real attack on \realsys is equivalent
to some ideal attack on \idealsys:
\[
   \realsys \emulates \idealsys \quad\defined\quad
   \forall \realatk.\; \exists \idealatk.\;
   \sembr{\realsys \mid \realatk} \squigeq \sembr{\idealsys \mid \idealatk}
\]

Here, \realatk ranges over allowed attackers on \emph{real programs}, \idealatk
    ranges over allowed \emph{ideal program} attackers, and the relation
    $\squigeq$ denotes \emph{behavioral equivalence} of program--adversary
    pairs.
\end{definition}

In other words, for every real attack \realatk, it must be possible to
construct an ideal attack \idealatk that \emph{simulates} it: the ideal attack
\idealatk is the \emph{simulator} for the real attack \realatk. Consequently,
the real--ideal paradigm is also known as \emph{simulation-based security}.

As stated in \cref{sub:Specifying and Enforcing Security Requirements}, the
notion of behavior $\sembr{\cdot}$ includes both system input--output (I/O) as well
as adversary leakage, which models information that the adversary learns. The
former captures \emph{integrity} requirements, while the
latter captures \emph{confidentiality} requirements.

We present an illustrative example for confidentiality. Suppose a
real system adversary \realatk can learn and leak the value of a
variable \vleak. However, the real system \realsys ensures that \vleak
is always equal to $42$. In this situation, \realsys emulates an ideal system
\idealsys that does not let its adversaries read any information; 
although \realatk can read \vleak and leak it, this
behavior can be simulated by an \idealatk that leaks $42$
instead of \vleak. On the other hand, if \realsys did store some
secret data in \vleak---a confidentiality violation---it would be impossible
to construct a real--ideal emulation. When \realatk leaks \vleak,
\idealatk cannot simulate it by leaking the same value because \vleak
contains secret data unknown to \idealatk.

\Cref{def:risim} is parameterized by a notion of behavioral equivalence:
the relation $\squigeq$, the precise instantiation of which
depends on the use case. For instance, the original formulation of the
real--ideal paradigm~\cite{gmw87,UC} defines this equivalence of behaviors as an
\emph{indistinguishability game} between the real and ideal worlds, where we can
choose among \emph{information-theoretic}, \emph{statistical}, or
\emph{computational} indistinguishability.

For this work, we use a simple version of information-theoretic (perfect)
indistinguishability: the semantic function $\sembr{\cdot}$ gives the trace of
events that occurs for each initial input. We make this more precise after
presenting our system model (\cref{par:Behavioral equivalence}).

\subsection{Single-program properties}\label{par:Single program properties}

\Cref{def:risim} defines $\emulates$ as a binary predicate over \realsys\
and \idealsys. It is also valuable to tweak this definition to be a unary
predicate over a single program \prog, by only restricting the collection of
ideal attackers. That is, if \ratkset is the set of real attackers, we define
another \iatkset which contains restricted attackers. Such an ideal attack
restriction defines a unary predicate over \prog.

\begin{definition}[Unary real--ideal emulation]
\label{def:unaryrisim}
A program \prog with a real attacker set \ratkset \emph{emulates} a weaker
\iatkset when every $\realatk \in \ratkset$ is equivalent to some
$\idealatk \in \iatkset$.
\[
   \forall \realatk \in \ratkset.\; \exists \idealatk \in \iatkset.\;
   \sembr{\prog \mid \realatk} \squigeq \sembr{\prog \mid \idealatk}
\]
\end{definition}
All security properties considered in this work are formulated as
unary real--ideal statements.

\section{Information flow security}\label{sec:Information flow security}

Information flow security seeks to defend against a powerful threat model.
Adversaries are given intimate read and write access to some part of the system
state. For example, adversaries are assumed to be able to read the state of all
public variables and freely modify values of untrusted variables at any point
during execution. Such adversaries are arguably more powerful than any
real-world adversary in a wide range of scenarios, from applications using
untrusted libraries to even distributed systems implemented by principals with
mutual distrust.

\SK{The paragraph below repeats intro}
Noninterference~\cite{GM82} is the best-known information flow
security property. Applied to both confidentiality and integrity, it
requires an adversary to be completely prevented both from learning
secret data and from corrupting trusted data. Noninterference is
simple, can be formulated extensionally, and is fairly straightforward
to enforce. However, the nuanced security requirements of realistic
systems typically violate noninterference. Additionally, while
noninterference for confidentiality and integrity can be factored
into two independent noninterference statements, realistic security
requirements intertwine these two aspects.

\subsection{Beyond noninterference: NMIF}
\label{sub:Beyond noninterference: NMIF}
\SK{Andrew says, "Do we want to say that this makes a good test case for the real--ideal approach". I don't know what this means}

There have been a few prior attempts to formulate useful security
properties that capture more realistic system security concerns. A well-known example is
the following family of properties: \emph{robust declassification} (RD),
\emph{transparent endorsement} (TE), and \emph{nonmalleable information flow}
(NMIF). They attempt to define secure violations of noninterference for
confidentiality (RD), integrity (TE), or both (NMIF).
\ACM{
    Mention others just to be more inclusive here? Like delimited
    release, relaxed noninterference, whatever the flow locks stuff
    enforces?
}

\paragraph{Robust declassification}\label{par:Robust declassification}

Robust declassification informally says that an active attacker cannot
learn more by interfering than a passive attacker (one that only
observes) could.  Equivalently, the attacker cannot exploit its power
to corrupt information to cause the program to declassify more (or
different) secret information than it would have leaked anyway.

The statement \lstinline|p := u * s| violates robust declassification. Here
\lstinline|p| is a public variable visible to the attacker, while \lstinline|s|
is secret and not visible. If the untrusted variable \lstinline|u| is initially zero,
the passive attacker does not learn \lstinline|s| from seeing
\lstinline|p|. But since \lstinline|u| is untrusted, the attacker controls its value;
by changing the value of \lstinline|u| to be non-zero, an active attacker learns the value of \lstinline|s|.

\paragraph{Transparent endorsement}
\label{par:Transparent endorsement}

Transparent endorsement informally says that a value can only be endorsed
(raised in integrity) if the principal supplying the value to be endorsed can
already read it. The endorsed value contains nothing that its supplier
didn't already know---hence ``transparent''. Equivalently, values that could have
been \emph{mauled} should not be endorsed.

Our example in \cref{sub:Making sense of nonmalleable information flow}
is actually an instance of a transparent endorsement violation. Bob supplies
\lstinline"bob_bid", whose value is endorsed despite it not being transparent
to Bob.

\paragraph{Nonmalleable information flow}
\label{par:Nonmalleable information flow}

Nonmalleable information flow is informally a mixture of RD and
TE. RD still enforces integrity noninterference while allowing confidentiality
downgrading, whereas TE enforces confidentiality noninterference while
allowing integrity downgrading. NMIF enforces controlled relaxation of both
confidentiality and integrity.

\paragraph{Limitations}\label{par:Limitations}

The original state-machine based formalization~\cite{zm01b} of robust
declassification elegantly formulates RD as a requirement that every
attacker learn at most as much as a passive version of itself which removes
all of its adversarial actions. Thus, every active attacker learns no more
than a passive attacker. 
Unfortunately, this formalism does not seem to generalize well in later works
that extend it for programming languages and in nonmalleable information
flow~\cite{nmifc,cecchetti-nmpl,msz06,zdance03}. These definitions are
\emph{intensional}: they require instrumenting the run-time semantics of programs 
with \emph{downgrade events}.
Additionally, the formalism has a complicated exclusion
of \emph{inept attackers}, defined as attackers who do not use
their ability fully.

\SK{Andrew says stealing our own thunder}
\paragraph{A real--ideal intuition}

All of the above properties seek to effectively eliminate an aspect of the
real adversary's powers. Nonmalleable information flow seeks to eliminate
an adversary's ability to maul data, while robust declassification takes
a step further and seeks to eliminate its entire ability to corrupt data.
Equivalently, we would like to reason about programs as if all adversaries
had these limitations.\ACM{
    This seems like the key insight, which is slightly buried in this
    paragraph. Maybe call it out more.
}
As we show in \cref{sub:Semantic characterization of
information flow properties}, the real--ideal paradigm makes these statements
precise. Such limited adversaries will be our \emph{ideal} adversaries;
we require that seemingly more powerful real adversaries can be
\emph{simulated} by these weaker adversaries.

\section{Semantic framework}\label{sec:Semantic framework} 

\begin{figure}
  \centering
     \begin{syntax}
    \groupleft{
      \abstractCategory[\setfnt{Var}]{\mvarfnt{x}, \mvarfnt{y}}
      \qquad
      \abstractCategory[\setfnt{Val}]{\mvarfnt{v}}
      \qquad
      \abstractCategory[\setfnt{Bool}]{\mvarfnt{b}}
    }
    \categoryFromSet[\setfnt{BinOp}]{\oplus}{\setfnt{Val} \to \setfnt{Val} \to \setfnt{Val}}
    \categoryFromSet[\setfnt{Test}]{\mvarfnt{t}}{\setfnt{Val} \to \setfnt{Bool}}
    \categoryFromSet[\setfnt{Store}]{\mvarfnt{\sigma}}{\setfnt{Var} \to \setfnt{Val}}
    \separate
    \category[\setfnt{ValExp}]{\mvarfnt{e_v}}
      \alternative{\mvarfnt{x} \oplus \mvarfnt{y}}
    \category[\setfnt{BoolExp}]{\mvarfnt{e_b}}
      \alternative{\mvarfnt{t}(\mvarfnt{x})}
    \category[\setfnt{Assignment}]{\mvarfnt{a}}
      \alternative{\mvarfnt{x} := \mvarfnt{e_v}}
    \category[\setfnt{Command}]{\mvarfnt{c}}
      \alternative{a}
      \alternative{\kwfnt{if}\ \mvarfnt{e_b}\ \kwfnt{then}\ \mvarfnt{p}\ \kwfnt{else}\ \mvarfnt{p}}
      \\
      \alternative{\kwfnt{while}\ \mvarfnt{e_b}\ \kwfnt{do}\ \mvarfnt{p}}
      \alternative{\kwfnt{output}~\mvarfnt{x}}
    \category[\setfnt{Program}]{\mvarfnt{p}}
      \alternative{\varepsilon}
      \alternative{\mvarfnt{c} \mathbin{;} \mvarfnt{p}}
  \end{syntax}
    \caption{\impthreat\ syntax}
   \label{fig:impthreat-syntax-nodg}
\end{figure}

\begin{figure}
  \centering
    \begin{syntax}
    \abstractCategory[\setfnt{State_{adv}}]{\mvarfnt{ast}}
    \separate
    \category[\setfnt{Cmd_{opt}}]{\mvarfnt{c^?}}
      \alternative{\epsilon}
      \alternative{\mvarfnt{c}}
    \separate
    \category[\setfnt{Cont_{sys}}]{\syscont}
      \alternative{\epsilon}
      \alternative{[\mvarfnt{c^?}] \,::\, \syscont}
      \alternative{[\mvarfnt{c^?}\, \advdevil]\,::\, \syscont}
    \separate
    \category[\setfnt{Config}]{\mvarfnt{\kappa}}
      \alternative{\configuration{\mvarfnt{\sigma}}{\mvarfnt{ast}}{\syscont}}
  \end{syntax}
   \caption{Run-time configurations of \impthreat}
  \label{fig:impthreatconfig}
\end{figure}

\begin{figure}
  \centering
  \begingroup
\newcommand{\cmdcfg}[2]{\langle #1,\ #2 \rangle}
\newcommand{\sembrH}[1]{{\sembr{#1}\mvarfnt{\sigma}}}

\noindent
\begin{minipage}[t]{0.48\linewidth}
\begin{judgedfunction}{\sembr{\mvarfnt{e_v}}\mvarfnt{\sigma} = \mvarfnt{v}}{\sembrH}
  \case{\mvarfnt{x} \oplus \mvarfnt{y}}
  \mvarfnt{\sigma}(\mvarfnt{x}) \oplus \mvarfnt{\sigma}(\mvarfnt{y})
\end{judgedfunction}
\end{minipage}\hfill
\begin{minipage}[t]{0.48\linewidth}
\begin{judgedfunction}{\sembr{\mvarfnt{e_b}}\mvarfnt{\sigma} = \mvarfnt{b}}{\sembrH}
  \case{\mvarfnt{t}(\mvarfnt{x})}
  \mvarfnt{t}(\mvarfnt{\sigma}(\mvarfnt{x}))
\end{judgedfunction}
\end{minipage}

\bigskip

\begin{judgedrules}{\cmdcfg{\mvarfnt{c}}{\mvarfnt{\sigma}} \stparr \mvarfnt{\sigma}'}
  \inferrule[\named{S-Assign}{store:assign}]
    {\sembr{\mvarfnt{e_v}}\mvarfnt{\sigma} = \mvarfnt{v}}
    {\cmdcfg{\mvarfnt{x} := \mvarfnt{e_v}}{\mvarfnt{\sigma}} \stparr \mvarfnt{\sigma}[\mvarfnt{x} \mapsto \mvarfnt{v}]}
\end{judgedrules}

\bigskip

\begin{judgedrules}{\cmdcfg{\mvarfnt{c}}{\mvarfnt{\sigma}} \cntarr \mvarfnt{p}}
  \inferrule[\named{C-If-T}{cont:ift}]
    {\sembr{\mvarfnt{e_b}}\mvarfnt{\sigma} = \kwfnt{true}}
    {\cmdcfg{\kwfnt{if}\ \mvarfnt{e_b}\ \kwfnt{then}\ \mvarfnt{p}_1\ \kwfnt{else}\ \mvarfnt{p}_2}{\mvarfnt{\sigma}} \cntarr \mvarfnt{p}_1}
  \and
  \inferrule[\named{C-If-F}{cont:iff}]
    {\sembr{\mvarfnt{e_b}}\mvarfnt{\sigma} = \kwfnt{false}}
    {\cmdcfg{\kwfnt{if}\ \mvarfnt{e_b}\ \kwfnt{then}\ \mvarfnt{p}_1\ \kwfnt{else}\ \mvarfnt{p}_2}{\mvarfnt{\sigma}} \cntarr \mvarfnt{p}_2}
  \\
  \inferrule[\named{C-While-T}{cont:whilet}]
    {\sembr{\mvarfnt{e_b}}\mvarfnt{\sigma} = \kwfnt{true}}
    {\cmdcfg{\kwfnt{while}\ \mvarfnt{e_b}\ \kwfnt{do}\ \mvarfnt{p}}{\mvarfnt{\sigma}} \cntarr \mvarfnt{p}\,;\,\kwfnt{while}\ \mvarfnt{e_b}\ \kwfnt{do}\ \mvarfnt{p}}
  \and
  \inferrule[\named{C-While-F}{cont:whilef}]
    {\sembr{\mvarfnt{e_b}}\mvarfnt{\sigma} = \kwfnt{false}}
    {\cmdcfg{\kwfnt{while}\ \mvarfnt{e_b}\ \kwfnt{do}\ \mvarfnt{p}}{\mvarfnt{\sigma}} \cntarr \epsilon}
\end{judgedrules}
\endgroup
   \caption{Selected stepping rules for \impthreat\ commands}
  \label{fig:impthreat-semantics}
\end{figure}

\begin{figure*}
  \centering
  \begingroup
\newcommand{\shadedcell}[1]{\tikz[baseline=(c.base)]{\node[fill=black!10, rounded corners=2pt,
    inner sep=2pt, minimum width=1.6em] (c) {#1};}}
\newcommand{\statetriple}[3]{\tikz[baseline=(o.base)]{\node[draw, rounded corners=3pt, inner sep=3pt] (o)
      {$\mvarfnt{v_p}$\shadedcell{#1}\,$\mvarfnt{v_s}$\shadedcell{#2}\,$\mvarfnt{v_u}$\shadedcell{#3}};}}

\newcommand{\examplecmd}{\tikz[baseline=(k.base)]{\node[fill=black!10, rounded corners=2pt,
    inner sep=3pt] (k) {$\mvarfnt{v_p} := \mvarfnt{v_s} * \mvarfnt{v_u}$};}}

\newcommand{\thinkingskull}[1]{\begin{tikzpicture}[baseline=(s.base)]
    \node[anchor=base] (s) at (0,0) {\advdevil[1]};
    \node[anchor=south west, draw, rounded corners=3pt, inner sep=3pt,
          minimum height=1.8em] (b)
      at (0.10,0.65) {#1};
    \draw ($(s.north)!0.55!(b.south west) + (0,-0.15)$) circle (0.045);
    \draw ($(s.north)!0.65!(b.south west) + (0.05,0)$)  circle (0.07);
  \end{tikzpicture}}

\scalebox{1.05}{\begin{tikzpicture}[
  font=\small,
  arr/.style={-{Stealth[length=2.5mm]}, thick},
  arrlbl/.style={midway, above, align=center, font=\scriptsize},
]

\node[anchor=west] (initst) at (0,0) {\textbf{Initial state}\quad\statetriple{0}{42}{0}};
\node[anchor=west] (cmdlbl) at (0,-0.8) {\textbf{Current command}\quad \examplecmd};

\node[anchor=north east, align=left, draw, rounded corners=3pt, inner sep=4pt]
  at (10,0.5)
  {$\mvarfnt{v_p}$ public\\
   $\mvarfnt{v_s}$ secret\\
   $\mvarfnt{v_u}$ untrusted};

\newcommand{\arrlen}{3.2}    

\newcommand{\nrow}[5]{\node[anchor=east] at (-0.2,#1) {#2};
  \node[anchor=west] at (0,#1)    {#3};
  \draw[arr] (\arrL,#1) -- node[arrlbl] {#4}
             ($(\arrL,#1)+(\arrlen,0)$);
  \node[anchor=west] at ($(\arrL,#1)+(\arrlen,0)$) {#5};
}

\def\arrL{3.0}    

\nrow{-2.0}{\circnum{1}}{\statetriple{0}{42}{0}}
  {\advdevil[0.6]\ corrupts $\mvarfnt{v_u}$ to $1$}{\statetriple{0}{42}{1}}
\nrow{-3.3}{\circnum{2}}{\statetriple{0}{42}{1}}
  {\examplecmd\ runs}{\statetriple{42}{42}{1}}

\nrow{-4.7}{\circnum{3}}{
\statetriple{42}{42}{1}
}
  {\advdevil[0.6]\ learns public variables}
  {\thinkingskull{``$\mvarfnt{v_p}$ is $42$''}}

\end{tikzpicture}}
\endgroup
   \caption{Illustration of command execution composed with adversary execution}
  \label{fig:compose-step}
\end{figure*}

\begin{figure}
  \centering
  \begingroup
\newcommand{\branchdiagram}{\begin{tikzpicture}[
arr/.style={-{Stealth[length=2.5mm]}, thick},
  cellbox/.style={draw, minimum width=2.6em, minimum height=2em, inner sep=2pt},
  restbox/.style={draw, pattern=north east lines, pattern color=black!40,
                  minimum width=3.5em, minimum height=2em},
]
\node[cellbox, anchor=west] (s0)  at (3, 0)    {$\mvarfnt{c}\,\advdevil$};
  \node[restbox, anchor=west] (s0r) at (s0.east) {$\mvarfnt{k'}$};
\node[cellbox, anchor=west] (t1) at (0,-1.9)   {$\mvarfnt{c}_1^t\,\advdevil$};
  \node[cellbox, anchor=west] (t2) at (t1.east)  {$\mvarfnt{c}_2^t\,\advdevil$};
  \node[cellbox, anchor=west] (t3) at (t2.east)  {$\mvarfnt{c}_3^t$};
  \node[restbox, anchor=west] (tr) at (t3.east)  {$\mvarfnt{k'}$};
\node[cellbox, anchor=west] (f1) at (4.0,-1.9) {$\mvarfnt{c}_1^f\,\advdevil$};
  \node[cellbox, anchor=west] (f2) at (f1.east)  {\advdevil};
  \node[restbox, anchor=west] (fr) at (f2.east)  {$\mvarfnt{k'}$};
\coordinate (jtop) at ($(s0.south)!0.5!(s0r.south)$);
  \coordinate (j)    at ($(jtop) + (0,-0.8)$);
  \draw (jtop) -- (j);
  \draw (j) -- (j -| t2.north)
    node[midway, above, font=\scriptsize] {$\llbracket\mvarfnt{b}\rrbracket = \mathit{true}$};
\coordinate (fmid) at ($(f1.north west)!0.5!(fr.north east)$);
  \draw (j) -- (j -| fmid)
    node[midway, above, font=\scriptsize] {$\llbracket\mvarfnt{b}\rrbracket = \mathit{false}$};
  \draw[arr] (j -| t2.north) -- (t2.north);
  \draw[arr] (j -| fmid) -- (fmid);
\end{tikzpicture}}

\newcommand{\condbox}{\scalebox{0.85}{\tikz[baseline=(ifb.base),
        cb/.style={draw, minimum width=2.4em, minimum height=1.8em, inner sep=2pt}]{\node[anchor=west]    (ifb)   at (0,0)                   {\textbf{if } $\mvarfnt{b}$ \textbf{then}};
    \node[cb, anchor=west](cct1)  at ($(ifb.east)+(0.2,0)$)  {$\mvarfnt{c}_1^t$};
    \node[cb, anchor=west](cct2)  at (cct1.east)             {$\mvarfnt{c}_2^t$};
    \node[cb, anchor=west](cct3)  at (cct2.east)             {$\mvarfnt{c}_3^t$};
    \node[anchor=west]    (elsel) at ($(cct3.east)+(0.2,0)$) {\textbf{else}};
    \node[cb, anchor=west](ccf1)  at ($(elsel.east)+(0.2,0)$) {$\mvarfnt{c}_1^f$};
    \node[draw, rounded corners=4pt, inner sep=3pt,
          fit=(ifb)(cct1)(cct2)(cct3)(elsel)(ccf1)] {};
  }}}

\begin{tikzpicture}[
  font=\small,
  arr/.style={-{Stealth[length=2.5mm]}, thick},
  cellbox/.style={draw, minimum width=2.6em, minimum height=2em, inner sep=2pt},
  restbox/.style={draw, pattern=north east lines, pattern color=black!40,
                  minimum width=3.5em, minimum height=2em},
]

\node[anchor=west] (cmdl) at (0, 2.2)
  {\textbf{Command} $\boldsymbol{\mvarfnt{c}}$\textbf{:}\quad\condbox};
\node[anchor=west] at (0, 1.4)
  {\textbf{Substep request from \advdevil[0.6]: $2$}};

\node[anchor=north west] at (0, 1) {\branchdiagram};

\end{tikzpicture}
\endgroup
   \caption{Zipping substep semantics}
  \label{fig:zip-step}
\end{figure}

\emph{Threat models}, which define the power of \emph{adversaries}
running concurrently with programs, are crucial for defining security
properties.  The security offered by a system depends on the power of
the adversary attacking it. Any system can defend against an adversary
that corrupts and observes nothing; no system can defend against an
omnipotent adversary.  The correctness of a defense is always defined
with respect to the threat model it was designed for.

Modeling threats in a language-based formalism requires careful thought. The
formalism needs to model stateful adversaries which are able to learn
information from and corrupt parts of program state. Additionally, the framework
needs to uniformly model not just strong real-world adversaries, but also
weaker specification-level adversaries.

Modeling strong adversaries is sometimes at odds with modeling weak ones. For
instance, a strong adversary should be able to run at all program execution
points. But a naive modeling of this power creates an unintended information
channel in the presence of control flow. In more detail, if a program's
\kwfnt{if}-command has branches of different lengths, then 
an adversary could learn which branch was chosen by counting the number of steps 
it takes. Our semantic framework captures
both strong and weak adversaries without baking in such implicit information
channels.

We model computation in the presence of an adversary using \impthreat, an
imperative while-language enriched with \emph{adversarial semantics}. Language
syntax (\cref{fig:impthreat-syntax-nodg}) is presented by defining the syntax
of \emph{commands} and \emph{programs} mutually inductively. \impthreat's
syntax is standard: it has assignments, conditionals (\kwfnt{if}), unbounded
loops (\kwfnt{while}), and \kwfnt{output} statements to generate observations.
Program state consists of a \emph{store}: a \emph{variable}--\emph{value} map.

In \impthreat, resources that need protection are variables. A program runs
along with an adversary \atk, which may interact with it. The limits of
the adversary's influence are defined using its \emph{adversarial power} (\cref
{def:advpower}).

\begin{definition}[Adversarial power]
\label{def:advpower}

\emph{Adversarial power} is a pair $\langle \setfnt{P}, \setfnt{U} \rangle$ where
$\setfnt{P}, \setfnt{U} \subseteq \setfnt{Var}$ are the \emph{public} and
\emph{untrusted} sets of variables, respectively.
\end{definition}

An adversary can \emph{read} public variables and \emph{corrupt} untrusted
variables. Not only can they corrupt by providing new values for variables,
they can \emph{maul} by supplying a \emph{transformation strategy} for
untrusted variables. The new values of mauled variables may depend on their old
values or on other variables. Modeling mauling is necessary as realistic
adversaries do have this ability. For instance, a UNIX process that has write
but no read permission for a file can selectively zero out any of its pages or
append arbitrary data. Another example comes from cryptography: with a
\emph{malleable} encryption scheme, adversaries can apply a known
transformation to an inaccessible secret message.

Formally, the adversary is defined as a deterministic state machine: a triple
consisting of \stadvset, its set of states; \initadv, its initial state
construction; and \stepadv, which describes how it steps. The initial state
construction allows the adversary to initialize its state from the public
variables of the initial program state. 

Adversary state transitions are  described by \stepadv: a pair $\langle \cmdpre, \cmdpost \rangle$,
where $\cmdpre \in \stadvset \to \storeset \to \storeset$ and $\cmdpost \in
\stadvset \to \storeset \to (\stadvset \times \natset)$. We use \cmdpre to
describe adversarial behavior \emph{before} a command runs; \cmdpre defines a
store transformation, modeling how the adversary causes corruption. We use
\cmdpost to describe adversarial behavior \emph{after} a command runs. Using
the current state and the public store, \cmdpost computes the next state as
well as the next \emph{substep count} (used to flexibly model control flow).
Additionally, \cmdpre and \cmdpost are restricted by \cref {def:advpower}:
\cmdpre can only maul untrusted variables, and \cmdpost can only read public
variables.

\impthreat's operational semantics are defined as a small-step relation over
system configurations (\cref {fig:impthreatconfig}). A configuration is a triple
$\configuration{\mvarfnt{\sigma}}{\mvarfnt{ast}}{\syscont}$ consisting of the
program store $\sigma$, the adversary's state \mvarfnt{ast} and the \emph{system
continuation} \syscont, which describes the rest of the computation. A system
continuation \syscont can be thought of as a generalization of a program;
whereas a program is just a list of commands, continuations are lists of
optional commands. A command may be marked by \advdevil to indicate whether
the adversary runs at that point. It is thus possible for the adversary and the
program to execute independently.

We define how a configuration steps by composing the stepping semantics for
commands with the adversary's next-step transformation.
\Cref{fig:impthreat-semantics} illustrates selected stepping rules for
commands. The relation $\stparr$ denotes how commands transform the program store, while
$\cntarr$ denotes new program branches that need to run. Trivial rules, like
\kwfnt{if}'s empty $\stparr$ behavior, are omitted.

These rules are \emph{sandwiched} between adversary execution when a system
continuation runs. More precisely, before a current command \mvarfnt{c} runs,
the adversary may corrupt any untrusted variables, and after \mvarfnt{c}
runs, the adversary computes its next state by possibly learning any
public variables. \Cref{fig:compose-step} shows an illustrative example.

Computing the next system continuation requires care as it might
bake in implicit information channels. To solve this problem, we introduce a
novel \emph{zipping substep semantics} (\cref {fig:zip-step}): at each step,
an adversary explicitly chooses \emph{any number of substeps} it wishes to run
for, regardless of the control path taken by the current command. 
For example, in an \kwfnt{if}-command, 
if the number of substeps chosen is less than the branch length, the adversary does
not run when the branch's excess commands do. If the number of substeps chosen
is greater, the adversary runs alone after the branch finishes. This new technique allows
us to handle secret and untrusted control flow, which was out of reach of prior
related work~\cite{viaduct-formal}, without weakening the adversary's ability to 
time execution by observing public events.

\paragraph{Extensionality versus threat models}\label{par:Extensionality versus threat models}
While adversaries interacting with internal program state at run time may seem
intensional, this is not so. This choice reflects the realistic \emph{threat
model} that is also shared with the original NMIF paper. Of course, other
weaker threat models are possible: for example, allowing adversaries to run only
at the start and end of execution~\cite{vsi96}. Such a restriction would weaken our results.

\subsection{Real--ideal emulation}
\label{sub:Real--ideal emulation}

Recall the definition of unary real--ideal emulation (\cref{def:unaryrisim}):
\[
   \forall \realatk \in \ratkset.\; \exists \idealatk \in \iatkset.\;
   \sembr{\prog \mid \realatk} \squigeq \sembr{\prog \mid \idealatk}
\]

In order to instantiate this definition for \impthreat, we need to define the following:
(1) the initial configuration $(\prog\; |\; \atk)$, 
(2) the notion of behavior $\sembr{\cdot}$, and
(3) behavioral equivalence ($\ba \squigeq \bb$).

\paragraph{The initial configuration}\label{par:The initial configuration}
We describe the starting configuration (\cref{fig:impthreatconfig})
for a given \impthreat program \prog, when run with \atk. The initial
configuration \initcfg is constructed from a given
\emph{input store} \siginp with the \mvarfnt{\sigma}
component of \initcfg  equal to \siginp.
The adversary computes its initial state using \mvarfnt{adv_{init}}. That
is, the \mvarfnt{ast} component of \mvarfnt{\kappa_{init}} is equal to
$\mvarfnt{adv_{init}}(\siginp)$. Finally, the system continuation
\mvarfnt{k}\ is constructed from the program \prog. In \mvarfnt{k}, the
adversary gets to run at every one of \prog's commands.

\paragraph{Defining behavior}\label{par:Defining behavior}
\begin{figure}
  \centering
  \begin{syntax}
  \abstractCategory[\setfnt{Leak}]{\mvarfnt{lk}}
  \separate
  \category[\setfnt{Event_{cmd}}]{\evc}
    \alternative{\epsilon}
    \alternative{\langle \mvarfnt{x}, \mvarfnt{v} \rangle}
  \category[\setfnt{Event_{adv}}]{\eva}
    \alternative{\epsilon}
    \alternative{\mvarfnt{lk}}
  \separate
  \category[\setfnt{Event}]{\mvarfnt{ev}}
    \alternative{\langle \evc, \eva \rangle}
    \alternative{\termev}
\end{syntax}
   \caption{\impthreat events}
  \label{fig:events}
\end{figure}

\begin{figure}
  \centering
  \begin{judgedrules}{\langle\mvarfnt{c}, \mvarfnt{\sigma}\rangle \evarr \mvarfnt{ev_c}}
  \inferrule[\named{Ev-Output-Trust}{ev:output-trust}]
    {\mvarfnt{x} \notin \setfnt{U}}
    {\langle \kwfnt{output}~\mvarfnt{x}, \mvarfnt{\sigma}\rangle \evarr \langle\mvarfnt{x}, \mvarfnt{\sigma}(\mvarfnt{x})\rangle}
  \and
  \inferrule[\named{Ev-Output-Untrust}{ev:output-untrust}]
    {\mvarfnt{x} \in \setfnt{U}}
    {\langle \kwfnt{output}~\mvarfnt{x}, \mvarfnt{\sigma}\rangle \evarr \epsilon}
\end{judgedrules}
   \caption{Selected command events}
  \label{fig:cmd-events}
\end{figure}

The behavior for a given program \prog when run with \atk is given by the
\emph{trace of events} produced by \initcfg for every input \siginp.
\Cref{fig:events} presents the syntax of an event. An event is either a pair
$(\evc, \eva)$, or a termination event (\termev). In the former, \evc represents
system I/O, and \eva is an adversarially \emph{leaked value}. 

A system I/O event \evc is a variable--value pair, and represents the output of
a variable \mvarfnt{x} along with its value. It only makes sense to consider
system I/O for \emph{honest} variables, as adversaries corrupt untrusted
variables arbitrarily. Formally, only \emph{trusted} variable outputs are
recorded. \Cref{fig:cmd-events} shows the precise semantics. In addition to
\evc, the adversary may leak an \eva, depending on its state. In general, an
adversary that knows more is able to leak more.

An event with both \evc and \eva equal to $\epsilon$ is a \emph{stutter}: a
silent step.

\paragraph{Behavioral equivalence}\label{par:Behavioral equivalence}
Our notion of behavioral equivalence builds on \emph{trace
equivalence}. Comparing traces requires ensuring \emph{stutter-invariance}; all
stutters are erased before comparison. Moreover, we make a distinction as to
whether termination behavior matters. \emph{Termination
sensitivity}~\cite{zm03,Askarov+:Termination} warrants \emph{strong trace
equivalence}, while \emph{termination insensitivity} requires only \emph{weak
trace equivalence}.

\begin{definition}[Strong trace equivalence]
\label{def:traceeqstr}
Traces \tracea and \traceb are \emph{strongly equal} (denoted
$\tracea\ssquigeqt\traceb $) if they are equal after stripping all stuttering
events.
\end{definition}

\begin{definition}[Weak trace equivalence]
\label{def:traceeqweak}
Traces \tracea and \traceb are \emph{weakly equal} (denoted
$\tracea\wsquigeqt\traceb $) if they are strongly equal, or if they have strongly
equal prefixes, after which one of them stutters forever.
\end{definition}

To rule out pathological ideal attackers that might halt system
execution by blocking progress---making the real and ideal trace trivially
weakly equal---we formulate an additional \emph{fairness restriction} on ideal
attackers.

\begin{definition}[Fair ideal attacker]
\label{def:fairidlatk}
An attacker \idealatk is \emph{fair} for a \prog and an attack \realatk if it
    never takes a step without one of \prog or \realatk also taking a step.
    Thus, a fair ideal attacker can never block progress alone.
\end{definition}

In this paper, this fairness restriction is required for all real--ideal
statements that use weak trace equivalence. Note that by making the ideal
attacker weaker, we \emph{strengthen} the security guarantee offered by
real--ideal emulation: the real attacker is shown equivalent in power to a
weaker attacker.

Behavioral equivalence $\sembr{\prog \mid \realatk} \squigeq \sembr{\prog \mid \idealatk}$ is then defined as
trace equivalence on every input store:
\[  
   \forall \siginp.\;
   \sembr{\prog \mid \realatk} (\siginp) \squigeqt \sembr{\prog \mid \idealatk} (\siginp)
\]

Again, equivalence can be strong (\cref{def:traceeqstr}) or weak
(\cref{def:traceeqweak}). We write $\ssquigeq$ when using $\ssquigeqt$ and
$\wsquigeq$ when using $\wsquigeqt$.

\subsection{Semantics of information flow properties}\label{sub:Semantic characterization of information flow properties}
Our formalization of \impthreat and its definition of real--ideal emulation
(\cref{sub:Real--ideal emulation}) give the necessary tools to formulate simple
and extensional security property definitions. Here, we describe how to
formalize the security properties described in \cref{sub:Beyond noninterference:
NMIF}: nonmalleable information flow, robust declassification,
transparent endorsement, and noninterference.

The definitions of all these security properties are presented as restrictions
on ideal adversary abilities. To express these restrictions, we require the
following building blocks.

\begin{definition}[Passive adversary]
\label{def:passiveadv}
An adversary is \emph{passive} when it does not use its corruption
power. While it may learn and thus leak the store's public view, it leaves the
store's untrusted view unchanged.
\end{definition}

\begin{definition}[Incurious adversary]
\label{def:incuriousadv}
An adversary is \emph{incurious} when it does not read anything from
the program store; its next state depends only on its current state.
\end{definition}

\begin{definition}[Nonmauling adversary]
\label{def:nonmaulingadv}
An adversary is \emph{nonmauling} when it corrupts untrusted variables 
only with values it supplies explicitly.
\end{definition}

\begin{table}
\centering 
\caption{Ideal adversary restrictions}
\label{tab:restr-table}
\begin{tabular}{|l|l|cc|}
\hline
Adversary     & Property      & Restrict\textsubscript{int} & Restrict\textsubscript{conf} \\ \hline
\niatkset     & NI            & Passive    & Incurious \\
\rdatkset     & RD            & Passive    & ---       \\
\teatkset     & TE            & Nonmauling & Incurious \\
\nmifatkset   & NMIF          & Nonmauling & ---       \\
\hline
\end{tabular}
\end{table}

From  these building blocks, we construct four sets of ideal adversaries \niatkset,
\rdatkset, \teatkset, and \nmifatkset, as summarized in \cref{tab:restr-table}.
For instance, an adversary in \teatkset is both nonmauling and incurious. Using
these adversaries, we can extensionally define the NMIF family of properties.

\begin{definition}[Nonmalleable information flow]
\label{def:nmif-def}
A program \prog satisfies nonmalleable information flow if every adversary
has an equivalent adversary that cannot maul.
\[
   \forall \realatk \in \ratkset.\; \exists \idealatk \in \nmifatkset.\;
   \sembr{\prog \mid \realatk} \squigeq \sembr{\prog \mid \idealatk}
\]

\end{definition}

\begin{definition}[Robust declassification]
\label{def:rd-def}
A program \prog satisfies robust declassification if every adversary has an
equivalent passive adversary.
\[
   \forall \realatk \in \ratkset.\; \exists \idealatk \in \rdatkset.\;
   \sembr{\prog \mid \realatk} \squigeq \sembr{\prog \mid \idealatk}
\]
\end{definition}

\begin{definition}[Transparent endorsement]
\label{def:te-def}
A program \prog satisfies transparent endorsement if every adversary has an
equivalent adversary that is both incurious and cannot maul.
\[
   \forall \realatk \in \ratkset.\; \exists \idealatk \in \teatkset.\;
   \sembr{\prog \mid \realatk} \squigeq \sembr{\prog \mid \idealatk}
\]
\end{definition}

\begin{definition}[Noninterference]
\label{def:ni-def}
A program \prog satisfies noninterference if every adversary has an
equivalent adversary that is both passive and incurious.
\[
   \forall \realatk \in \ratkset.\; \exists \idealatk \in \niatkset.\;
   \sembr{\prog \mid \realatk} \squigeq \sembr{\prog \mid \idealatk}
\]
\end{definition}

\subsection{Extensional versus intensional NMIF}\label{sub:Extensional versus intensional NMIF}
It is instructive to compare the original formulation of NMIF~\cite {nmifc}
with ours. The former is tied to a particular IFC type system for a functional
calculus. Its definitions rely on instrumenting the operational semantics
with events caused by syntactic downgrading expressions, and are thus
intensional. Unfortunately, its tight coupling to the IFC calculus
gets in the way of a meaningful mathematical comparison with our purely
semantic characterization.

However, the benefits of our extensional framework can be seen by further
analyzing the auction example~(\cref{fig:auction}), which the original
definition deems to violate NMIF. The real--ideal definition agrees, and
considers the example insecure because there is no equivalent nonmauling ideal
Bob for an adversarial Bob, who mauls using the transformation
\lstinline|bob_bid := alice_bid + 1|. 

Now, suppose that \lstinline|auctioneer| were modified so that the function
always returns 0. In this case, the code \emph{would satisfy} our extensional
definition of NMIF because a mauler cannot affect a constant function. But the
original intensional definition considers the modified code insecure because of
endorse events, even though no observer can distinguish
\lstinline|auctioneer| from a
constant function that always returns 0.

Beyond the increased accuracy of our extensional definition,
another important difference is that the original
definition requires a complex restriction to exclude \emph{inept attackers} who
do not use all their adversarial abilities. In our framework, inept real
attackers need no special treatment: they can be simulated by similarly
inept ideal attackers.\ACM{
    Maybe just cut the rest of the paragraph for length or save it for
    the TR?
}
For example, consider this program adapted from the
original paper: \lstinline"if s then p := u1 else p := 1 - u2", where
\lstinline"s" is secret, \lstinline"p" is public, and the \lstinline"u"
variables are untrusted and initially zero. This program satisfies robust
declassification according to both definitions. In our new definition, the
passive ideal attacker learns the secret \lstinline"s" and can thus simulate
any active real attacker. The common-case version of the original
4-hyperproperty definition requires the output value of \lstinline"p" to be
independent of the values of the \lstinline"u" variables. This is mostly true
\emph{except for the edge case} where $\texttt{\small u1} + \texttt{\small u2}
= 1$---an inept attack that learns nothing. There, ruling out such `irrelevant' inputs
requires additional, complex conditions that our framework does not need.

\subsection{Heterogeneous security}\label{sub:Heterogeneous Security}

Our discussion so far has fixed a single \emph{adversarial power}.
While this model might be acceptable for reasoning about single-user
programs, it limits the ability to faithfully model decentralized
systems with mutually distrusting principals.

In order to facilitate such reasoning, we generalize our real--ideal formalism
(\cref{def:risim}) to quantify over all reasonable adversarial powers. 
We use \advpower to denote such adversarial powers.

Formally, in the unary case, we require: 
\[
  \forall \realatk \in
  \advpower\text{-}\ratkset.\; \exists \idealatk \in \advpower\text{-}\iatkset.
  \sembr{\prog \mid \realatk} \squigeq \sembr{\prog \mid \idealatk}
\]
\noindent
That is, for a real attack at some particular level, there must exist an ideal
attacker at \emph{the same level}. For instance, a corrupted real Bob in an
Alice--Bob protocol must be simulated by a corrupted ideal \emph{Bob}, not some
other principal.  Our type system (\cref{sec:heterogeneous NMIFC}) offers
such expressive heterogeneous security guarantees.

\section{Soundness for practical IFC type systems}\label{sec:Real--ideal soundness for information flow types}
\begin{figure}
  \centering
  \begin{syntax}
  \category[\setfnt{Assignment}]{\mvarfnt{a}}
    \alternative{\ldots}
    \\
    \alternative{\mvarfnt{x} := \cmdendorse(\mvarfnt{y})}
    \\
    \alternative{\mvarfnt{x} := \cmddeclassify(\mvarfnt{y})}
\end{syntax}
   \caption{Extending \impthreat\ syntax with downgrading}
  \label{fig:impthreat-syntax-dg}
\end{figure}

Practical information flow type systems with
downgrading~\cite{myers-popl99,ps03,Paragon} have their syntax extended with
\cmdendorse and \cmddeclassify statements, which weaken noninterference (for
integrity and confidentiality respectively). Programmers audit these statements
to look for security vulnerabilities.  Using the real--ideal paradigm, we now
formalize this idea that such auditing suffices, as a soundness statement for
practical IFC type systems.

First, we extend \impthreat syntax with downgrading statements
(\cref{fig:impthreat-syntax-dg}), which are semantically assignments (with
similar $\stparr$, $\cntarr$, and $\evarr$ rules). Then, we formalize adversary
interaction only with downgrading statements by defining
\emph{downgrade-restricted} adversaries.

\begin{definition}[Downgrade-restricted adversary]
\label{def:dgonlyadv}
An adversary is \emph{downgrade-restricted} when it may only corrupt (without mauling)
the \emph{source variable} of a currently running nontrivial \cmdendorse statement,
and may only read the \emph{destination variable} of a currently running nontrivial
\cmddeclassify statement. 
\end{definition}

A downgrade statement that is noninterfering is considered trivial,
as it is not really a downgrade. 
The restriction
disallowing interactions with such statements reflects this. 
Note that this
restriction only \emph{strengthens} our security statement by shrinking the set of
ideal adversaries.
In practice, when giving heterogeneous security guarantees,
some downgrades may be noninterfering for some attackers; it is useful to not
have to reason about those downgrades.

\begin{definition}[Downgrade-restricted real--ideal emulation]
\label{def:dgonlyri}
Program \prog satisfies \emph{downgrade-restricted real--ideal emulation} if
    every attack \realatk is equivalent to a downgrade-restricted attack
    \idealatk:
\[
   \forall \realatk \in \ratkset.\; \exists \idealatk \in \dgatkset.\;
   \sembr{\prog \mid \realatk} \squigeq \sembr{\prog \mid \idealatk}
\]
\end{definition}

\subsection{Implications}\label{sub:Implications}
\Cref{def:dgonlyri} restricts system behavior to an abstract realm with
weakened adversaries. We now show the payoff of this abstraction: enforcing the
security properties defined in \cref{sub:Semantic characterization of information flow properties} 
for programs that satisfy \cref{def:dgonlyri}.

\SK{Emphasize that these are proofs}
NMIF (\cref{def:nmif-def}) immediately follows from \cref{def:dgonlyri}, which
constrains ideal attackers even more than NMIF does! The latter only allows
nonmauling corruption at \cmdendorse statements, which is stricter than
arbitrary nonmauling corruption. RD (\cref{def:rd-def}), TE
(\cref{def:te-def}), and NI (\cref{def:ni-def})  are enforced by placing
appropriate restrictions on program \cmdendorse and \cmddeclassify statements.
The only way for an \idealatk to cause corruption is via \cmdendorse
statements, so their absence implies RD. Similarly, the absence of
\cmddeclassify statements implies TE, because \cmddeclassify statements are the
only way for an \idealatk to learn information. Finally, the absence of both
\cmdendorse and \cmddeclassify statements implies NI, because an \idealatk now
has no power. We formally prove all the above results.

\section{Simplified NMIFC}\label{sec:Simplified NMIFC}

\begin{figure*}
  \centering

\noindent
\begin{minipage}[t]{0.65\linewidth}
\vspace{0pt}
\begin{judgedrules}{\judgec}
  \inferrule[\named{ST-Assign}{type:assign}]
    {\srcfn(\mvarfnt{a}) \flowto {\dstfn(\mvarfnt{a})} \\
     \mvarfnt{pc} \flowto \dstfn(\mvarfnt{a})}
    {\mvarfnt{pc} \wts \mvarfnt{a}}
  \and
  \inferrule[\named{ST-Endorse}{type:endorse}]
    {\advpowerof{\mvarfnt{y}} = \PU \\ \advpowerof{\mvarfnt{x}} = \PT}
    {\PT \wts \mvarfnt{x} := \kwfnt{endorse}(\mvarfnt{y})}
  \and
  \inferrule[\named{ST-Declassify}{type:declassify}]
    {\advpowerof{\mvarfnt{y}} = \ST \\ \advpowerof{\mvarfnt{x}} = \PT}
    {\PT \wts \mvarfnt{x} := \kwfnt{declassify}(\mvarfnt{y})}
  \\
  \inferrule[\named{ST-Output}{type:output}]
    {\mvarfnt{pc} \flowto \advpowerof{\mvarfnt{x}}}
    {\mvarfnt{pc} \wts \kwfnt{output}~\mvarfnt{x}}
  \and
  \inferrule[\named{ST-If}{type:if}]
    {\advpowerof{\mvarfnt{x}} \lbljoin \mvarfnt{pc} \wts \mvarfnt{p}_1 \\
     \advpowerof{\mvarfnt{x}} \lbljoin \mvarfnt{pc} \wts \mvarfnt{p}_2}
    {\mvarfnt{pc} \wts \kwfnt{if}\ \mvarfnt{t}(\mvarfnt{x})\ \kwfnt{then}\ \mvarfnt{p}_1\ \kwfnt{else}\ \mvarfnt{p}_2}
  \and
  \inferrule[\named{ST-While}{type:while}]
    {\advpowerof{\mvarfnt{x}} \lbljoin \mvarfnt{pc} \wts \mvarfnt{p}}
    {\mvarfnt{pc} \wts \kwfnt{while}\ \mvarfnt{t}(\mvarfnt{x})\ \kwfnt{do}\ \mvarfnt{p}}
\end{judgedrules}
\end{minipage}\hfill
\begin{minipage}[t]{0.32\linewidth}
\vspace{0pt}
\begin{judgedrules}{\judgep}
  \inferrule[\named{ST-Empty}{type:empty}]
    { }
    {\mvarfnt{pc} \wts \epsilon}
  \and
  \inferrule[\named{ST-Seq}{type:seq}]
    {\mvarfnt{pc} \wts \mvarfnt{c} \\ \mvarfnt{pc} \wts \mvarfnt{p}}
    {\mvarfnt{pc} \wts \mvarfnt{c} \mathbin{;} \mvarfnt{p}}
\end{judgedrules}

\bigskip

\[  \advpowerof{\cdot},\ \mvarfnt{pc} \in \{\, \PT, \PU, \ST, \SU \,\}  \]
\begin{align*}
    \srcfn(\mvarfnt{x} := \mvarfnt{y} \oplus \mvarfnt{z}) & = \advpowerof{y} \lbljoin  \advpowerof{z} \\
    \srcfn(\mvarfnt{x} := \cmdendorse(\mvarfnt{y})) &= \advpowerof{y}\\
    \srcfn(\mvarfnt{x} := \cmddeclassify(\mvarfnt{y})) &= \advpowerof{y} \\
    \dstfn(\mvarfnt{x} := \dots) &= \advpowerof{x}  \\
\end{align*}
\end{minipage}
   \caption{Four-point NMIFC typing rules for \emph{adversarial power} \advpower}
  \label{fig:nmifc-types}
\end{figure*}
We enforce \cref{def:dgonlyri} using a type system based on the 
NMIFC type system of \citet{nmifc}. Before presenting our full NMIFC type system
(\cref{fig:nmifc-types-het}), we elucidate core ideas with a simplified
version (\cref{fig:nmifc-types}). The simplification is that we fix an
adversarial power (\cref{def:advpower}) \advpower, which partitions the set of
variables into four classes: $\{\PT, \PU, \ST, \SU\}$, which stand for
\emph{public--trusted}, \emph{public--untrusted}, \emph{secret--trusted}, and
\emph{secret--untrusted}, respectively. As explained in \cref{def:advpower}, a
\emph{public} variable is readable by the adversary, while an \emph{untrusted}
variable can be corrupted by mauling.

Following standard information flow literature, these four classes have a
\emph{lattice} structure~\cite{denning-lattice,MITRE,integrity}, with the
binary operators \emph{meet} (\lblmeet) and \emph{join} (\lbljoin), and a
binary \emph{flows to} (\flowto) relation. Trusted variables can flow to
untrusted variables, and public variables can flow to secret variables. We can
thus define \flowto for these four classes component-wise. 

\Cref{fig:nmifc-types} is a fairly standard four-point static information flow
type system. It defines two mutually inductive typing judgments \judgec and
\judgep, which can be read as saying, ``the program \mvarfnt{p} / command
\mvarfnt{c} is well typed in context \mvarfnt{pc}''. Here \mvarfnt{pc}
represents influences on the current \emph{control-flow}, and constrains
control-flow statements (\cref{type:if,type:while}). Initially, the program
\mvarfnt{p} is public and trusted, and we define $\wts \prog$ to be $\PT \wts
\prog$.

\Cref{type:declassify} and \cref{type:endorse} are the main rules constraining
downgrading. They are the only rules that break noninterference,  and the only
program locations where the soundness statement (\cref{def:dgonlyri}) allows
ideal-adversary influence. Trivial noninterfering downgrades are also
allowed by \cref{type:assign}.

While this is a type system that only states static requirements for a
particular adversarial power \advpower, it captures all required key ideas. In
fact, our soundness proof for the full heterogeneous type system builds directly 
on top of this.

\subsection{Real--ideal soundness}\label{sub:risoundness simple}

The soundness theorem for our type system is a real--ideal statement that
constrains ideal adversaries to only downgrading interactions
(\cref{def:dgonlyri}). As \impthreat has \kwfnt{while} loops whose termination
behavior cannot be determined statically, our notion of trace equivalence is
\emph{weak} (\cref{def:traceeqweak}).

\begin{theorem}[Real--ideal soundness for simple NMIFC]
\label{thm:simplenmifcsound}
A well-typed program \prog satisfies \cref{def:dgonlyri}.
\begin{align*}
\wts \prog \implies &\forall \realatk \in \ratkset. \\
    &\exists \idealatk \in \dgatkset.~\sembr{\prog \mid \realatk} \wsquigeq \sembr{\prog \mid \idealatk}
\end{align*}
\end{theorem}

\paragraph{Intuitive argument}
\label{par:Intuitive argument}

Suppose we have a \prog such that $\wts \prog$. The type system
(\cref{fig:nmifc-types}) ensures \emph{noninterference} for most commands,
which enforces strict constraints on the integrity of trusted variables and
the privacy of secret variables. In this case, anything that the real attacker
\realatk can leak by reading public variables at any point in the program
can be \emph{simulated} and leaked by an \idealatk, by just computing from
\siginp's initial public input. Additionally, as only trusted events appear on
the traces, \idealatk's lack of corruption power does not hurt us, and we can
ensure that $\sembr{\prog \mid \realatk} \wsquigeq \sembr{\prog \mid
\idealatk}$.

\paragraph{Challenges}
\label{par:Challenges}

The intuitive argument sweeps several tricky details under the rug. Unlike
traditional program refinement proofs, which often just involve some sort of
bisimulation between two programs, this proof is more akin to a \emph{compiler
correctness} proof. We \emph{compile} the real attack \realatk and program
\prog to explicitly construct the \emph{simulator}, which we denote 
\simatk. Although \simatk has to ``just simulate'' \realatk's leakage behavior,
this requires that it can independently compute the values of variables that
\realatk leaks, which in turn requires that it keep in its memory a version
of the \emph{system continuation} that will run.

After the \simatk construction, showing the equivalence $\wsquigeq$ requires the
construction of another \emph{bisimulation relation} relating the real and ideal
worlds. This is nontrivial, for we are dealing with four concurrently executing
processes: the real program, the real attacker, the ideal program, and the ideal
attacker. We need to show that this bisimulation relation is preserved as the
two systems execute. 
In order to tame this complexity, we join the two systems into a combined configuration,
which lets us reduce bisimulation to a
unary problem of invariant preservation.
We also need to show that $\simatk \in \dgatkset$, while additionally ensuring
that \simatk is \emph{fair} (\cref{def:fairidlatk}).

\paragraph{Proof strategy}
\label{par:Proof strategy}
\Cref{sec:NMIF soundness} presents the proof sketch, and
has the following components: (1) simulator construction (\cref{sub:The
simulator construction}), (2) proving behavioral equivalence using bisimulation
(\cref{sub:Proving eq,sub:Combined configurations,sub:Extraction
commutativity,sub:The bisimulation relation}), (3) demonstrating that the
simulator is in \dgatkset (\cref{sub:The simulator is downgrade-restricted}),
and (4) showing fairness for our simulator construction (\cref{sub:Simulator
fairness}).

\section{Heterogeneous NMIFC}\label{sec:heterogeneous NMIFC}
\begin{figure*}
  \centering
  
\begin{judgedrules}{\pc \wt \cmd}
  \inferrule[\named{T-Assign}{type:assign-het}]
    {\lblof{\mvarfnt{x}} \flowto \lblof{\mvarfnt{z}} \\
     \lblof{\mvarfnt{y}} \flowto \lblof{\mvarfnt{z}} \\
     \mvarfnt{pc} \flowto \lblof{\mvarfnt{z}}}
    {\mvarfnt{pc} \wt \mvarfnt{z} := \mvarfnt{x} \oplus \mvarfnt{y}}
  \and
  \inferrule[\named{T-Endorse}{type:endorse-het}]
    {\conf{\lblof{\mvarfnt{y}}} = \conf{\lblof{\mvarfnt{x}}} \\
     \integ{\lblof{\mvarfnt{x}}} \stronger \integ{\lblof{\mvarfnt{y}}} \\
     \mvarfnt{pc} \flowto \lblof{\mvarfnt{x}} \\
     \uncomp{\lblof{\mvarfnt{y}} \lbljoin \mvarfnt{pc}}}
    {\mvarfnt{pc} \wt \mvarfnt{x} := \kwfnt{endorse}(\mvarfnt{y})}
  \and
  \inferrule[\named{T-Declassify}{type:declassify-het}]
    {\integ{\lblof{\mvarfnt{y}}} = \integ{\lblof{\mvarfnt{x}}} \\
     \conf{\lblof{\mvarfnt{y}}} \stronger \conf{\lblof{\mvarfnt{x}}} \\
     \mvarfnt{pc} \flowto \lblof{\mvarfnt{x}} \\
     \uncomp{\lblof{\mvarfnt{y}} \lbljoin \mvarfnt{pc}}}
    {\mvarfnt{pc} \wt \mvarfnt{x} := \kwfnt{declassify}(\mvarfnt{y})}
  \\
  \inferrule[\named{T-Output}{type:output-het}]
    {\mvarfnt{pc} \flowto \lblof{\mvarfnt{x}}}
    {\mvarfnt{pc} \wt \kwfnt{output}~\mvarfnt{x}}
  \and
  \inferrule[\named{T-If}{type:if-het}]
    {\lblof{\mvarfnt{x}} \lbljoin \mvarfnt{pc} \wt \mvarfnt{p_1} \\
     \lblof{\mvarfnt{x}} \lbljoin \mvarfnt{pc} \wt \mvarfnt{p_2}}
    {\mvarfnt{pc} \wt \kwfnt{if}\ \mvarfnt{t}(\mvarfnt{x})\ \kwfnt{then}\ \mvarfnt{p_1}\ \kwfnt{else}\ \mvarfnt{p_2}}
  \and
  \inferrule[\named{T-While}{type:while-het}]
    {\lblof{\mvarfnt{x}} \lbljoin \mvarfnt{pc} \wt \mvarfnt{p}}
    {\mvarfnt{pc} \wt \kwfnt{while}\ \mvarfnt{t}(\mvarfnt{x})\ \kwfnt{do}\ \mvarfnt{p}}
\end{judgedrules}

\bigskip

\begin{judgedrules}{\pc \wt \prog}
  \inferrule[\named{T-Empty}{type:empty-het}]
    { }
    {\mvarfnt{pc} \wt \epsilon}
  \and
  \inferrule[\named{T-Seq}{type:seq-het}]
    {\mvarfnt{pc} \wt \mvarfnt{c} \\ \mvarfnt{pc} \wt \mvarfnt{p}}
    {\mvarfnt{pc} \wt \mvarfnt{c} \mathbin{;} \mvarfnt{p}}
\end{judgedrules}
   \caption{Heterogeneous NMIFC typing rules}
  \label{fig:nmifc-types-het}
\end{figure*}
\Cref{fig:nmifc-types} describes the static obligations a program must satisfy
to enjoy our real--ideal emulation (\cref{def:dgonlyri}), but only against one
\emph{particular} adversary \advpower. In decentralized systems, however,
defending against such a single adversarial power does not suffice; a secure
system must provide guarantees across \emph{all reasonable} adversarial
scenarios.

In order to achieve this, we first need a way to describe different adversarial
scenarios. Following standard practice, we do this using \emph{security labels}
drawn from a lattice of \emph{principals} \Pset~\cite{ml-ifc-97,flam}.
Principals in \Pset are ordered by the \emph{authority relation} \stronger.
When $\mvarfnt{p_1} \stronger \mvarfnt {p_2}$, we say that \mvarfnt{p_1}
\emph{acts for} \mvarfnt{p_2}: intuitively \mvarfnt{p_1} can do everything that
\mvarfnt{p_2} can. Finally, we assume that \Pset is \emph{bounded}: $\top$ is the
principal with the most authority and $\bot$ is the principal with the least
authority.

Variables are annotated with \emph{labels}. A
label \lbl is a \emph{pair} of principals denoting minimal \emph{read} and
\emph{write} requirements for data. For a label \lbl, we write $\integ{\lbl}$
for its write requirements, and $\conf{\lbl}$ for its read requirements. Read
and write permissions in UNIX are a good real-world example of what such
labels can encode. We write $\lblof{\mvarfnt{x}}$ for the label of a variable
\mvarfnt{x}.

In this setup, an adversarial power is also a label \advpower, which
partitions variables into the four classes $\{\PT, \PU, \ST, \SU\}$ in the
obvious way. If \advpower has the authority to read a variable, it is public;
if it has the authority to write a variable, it is untrusted. Additionally, we
assume \advpower to be \emph{reasonable}.

\begin{definition}[Reasonable adversarial power]
\label{def:reasonable-adv}
An adversarial power \advpower is \emph{reasonable} when $\conf{\advpower}
\stronger \integ{\advpower}$: the adversary's read power is at least as strong
as its write power; further, it cannot control $\top$.
\end{definition}

Intuitively, a reasonable adversary can only corrupt what it can also observe.
This assumption is standard~\cite{nmifc,viaduct-pldi21,zsm19,ifc-delegation}, and suffices to encode
the well-known \emph{malicious} and \emph{honest-but-curious} models from
cryptography. The absurd adversary that cannot read its own data is excluded.

In \cref{fig:nmifc-types-het}, we show our full NMIFC type system for a given
principal lattice \Pset and a variable-labeling function $\lblof{(\cdot)}$. Like
the simplified system (\cref{fig:nmifc-types}), it defines $\pc \wt \cmd$ and
$\pc \wt \prog$ mutually inductively, and has similar rules when there is
no downgrading.

\Cref{type:endorse-het,type:declassify-het} generalize
the simpler \cref{type:endorse,type:declassify}: rather than fixing destinations to be $\PT$ in a
$\PT$ context, their labels have to be \emph{uncompromised}.

\begin{definition}[Uncompromised label]
\label{def:uncompromised}
A label \lbl is \emph{uncompromised}, denoted $\uncomp{\lbl}$,
when $\integ{\lbl} \stronger \conf{\lbl}$: the integrity
component is at least as strong as the confidentiality
component~\cite{zsm19}.
\[
\uncomp{\lbl} \defined \integ{\lbl} \stronger \conf{\lbl}
\]
\end{definition}

A principal that can write to an $\uncomp {\lbl}$ variable can also read it.

We now prove that a well-typed NMIFC program \emph{projects} onto its well-typed
four-point counterpart. We write $\wt \prog$ to mean $\pc \wt \prog$, where $\integ
{\pc} = \top$, and $\conf {\pc} = \bot$. That is, it is well-typed in the
\emph{most} public and trusted context.

\begin{lemma}[NMIFC projection]
\label{lemma:nmifc-project}
For all reasonable attackers, a well-typed program \prog remains well-typed in the
four-point case.
\[
    \forall \advpower.\; \fnfnt{reasonable} (\advpower)\;\land \wt \prog \implies \wts \prog
\]
\end{lemma}

The key idea in this proof is to recognize that nearly every typing
rule in \cref{fig:nmifc-types-het} maps exactly to its corresponding
four-point rule in \cref{fig:nmifc-types}. A minor exception is the pair
of \cref{type:endorse-het,type:declassify-het}. In addition to mapping onto
\cref{type:endorse,type:declassify}, they might be noninterfering and thus also
map onto \cref{type:assign}.

We now prove our final heterogeneous real--ideal soundness theorem.

\begin{theorem}[NMIFC heterogeneous real--ideal soundness]
\label{theorem:nmifc-het-ri}
A well-typed program \prog satisfies \cref{def:dgonlyri} for all reasonable adversarial powers \advpower. 
\begin{align*}
    \wt \prog \implies  & \forall \realatk \in \advpower\text{-}\ratkset. \\
                        & \exists \idealatk \in \advpower\text{-}\dgatkset.\;
\sembr{\prog \mid \realatk} \wsquigeq \sembr{\prog \mid \idealatk}
\end{align*}
\end{theorem}

\subsection{Connection to Viaduct}\label{sub:Connection to Viaduct}
This type system is a generalization of the type system used in the
Viaduct~\cite{viaduct-pldi21, viaduct-formal} compiler: in Viaduct, only public
and trusted control flow is permitted. This
work sheds light on the precise role that NMIFC plays in Viaduct. Its source
language only allows adversarial access to downgrading (like
\cref{def:dgonlyadv}), and Viaduct is able to connect this source threat model
to a more powerful real-world adversary because of \cref{theorem:nmifc-het-ri}
that NMIFC satisfies. We believe that this work should enable extending Viaduct
to handle more language features. 

\section{Beyond NMIF}
\label{sec:Beyond NMIF}
The NMIF family of properties was chosen as the primary case study for our
framework because, as less-understood 4-safety
hyperproperties, they present a challenge for any security
specification technique. Further, while they have been useful for
building secure systems and compilers~\cite{zznm02,jfabric,zsm19,viaduct-pldi21},
a simple and intuitive theoretical underpinning has been lacking.
Of course, the real--ideal framework is not tied to NMIF; its
generality permits other well-known security properties to be expressed
in the real--ideal paradigm, as we now discuss.
\ACM{
    This is all great. I did a little editing. The reviewer who wants
    to reject says, "So you did a deep evaluation on security property
    for which you cannot directly compare the semantic security
    condition, and you did a shallow evaluation on the security
    conditions for which a direct comparision is feasible?"
}

\paragraph{Trace refinement}\label{par:Trace refinement as ri}
Behavioral equivalence requires equivalence of both system I/O and adversarial
leakage (\cref{par:Behavioral equivalence}).
Considering only system I/O yields the well-known
notion of \emph{trace refinement}. A trace refinement specification is
nondeterministic and specifies all allowed executions; a conforming
implementation must ensure that all its own executions are allowed by the
specification. The real--ideal paradigm captures this nondeterminism as
\emph{adversarial influence}. As execution proceeds, adversarial influence
may cause different program behavior, which is precisely (demonic) nondeterminism!
Requiring that an ideal adversary exist for every real adversary is then
exactly equivalent to the refinement requirement. In this way, the real--ideal
paradigm satisfactorily explains demonic
nondeterminism~\cite{Dijkstra76a}. 

\paragraph{Relaxed noninterference}\label{par:Relaxed noninterference as ri}
Relaxed noninterference~\cite{lz05} generalizes confidentiality labels to
arbitrary computation representing a variable's secrecy policy. For instance, a
policy allowing the last 4 digits of a credit card number to be public could be
written as $\mvarfnt{policy} \defined \lambda i.\, i \bmod 10000$, and
input variable \mvarfnt{card\_num} would be annotated with \mvarfnt{policy}.
The goal then is to ensure that the information that an output public variable
gains in a program is restricted by this policy.

Relaxed noninterference maps directly into the real--ideal paradigm.
The ideal adversary only learns the information that a variable policy
allows at program initialization. Relaxed noninterference is then simply
real--ideal emulation against a real adversary that can read the values of
public output variables.

Finally, as relaxed noninterference generalizes delimited release~\cite{sm04}
and incorporates aspects of what-declassification~\cite{ss05}, it
follows that our framework can too; in fact, it can do more, as it is not
restricted to only confidentiality policies.

\paragraph{Disjunctive information flow}\label{par:Disjunctive information flow as ri}
Disjunctive information flow captures requirements such as separation of duty and
secure secret sharing~\cite{Chinese-Wall}.
Lattices must be generalized to
quantales~\cite{qoi}---structures with two kinds of ``join'' operators: $\vee$
and $\otimes$. $\text{Alice} \vee \text{Bob}$ denotes the common authority of
both Alice and Bob, while $\text{Alice} \otimes \text{Bob}$ denotes controlling
\emph{exactly one} of Alice or Bob.  

In a real--ideal setting, such policies are naturally expressed: the new set of
ideal attackers is simply the \emph{set union} of the component attacker sets.
As an illustrative example, consider secure secret sharing with two variables:
\lstinline{alice_secret} and \lstinline{bob_secret}, where an attacker is
allowed to learn at most one of the two. If \aliceatkset and \bobatkset are the ideal
attacker sets that are only allowed access to \lstinline{alice_secret} and
\lstinline{bob_secret} respectively, secret-sharing security now is real--ideal
emulation with the ideal attacker set $\aliceatkset \cup \bobatkset$.  This
definition, as well as examples that satisfy and do not satisfy it, are included
in our formalization.

\section{Evaluation and proof mechanization}
We formalize and prove all our theorems using the Rocq~\cite{rocq910} proof
assistant. This artifact includes a formally verified type checker for NMIFC, which we
use to type-check all examples from the original paper. As described in
\cref{sub:Proving eq}, proving soundness for NMIFC needs a fairly involved
bisimulation proof; we believe that this proof is the first formally
verified real--ideal emulation result for a core programming language calculus
with realistic security concerns that combine confidentiality and integrity.
Our total development totals about \textasciitilde5K lines of code. The proofs
have no admits, and only use the axiom for function extensionality. The longest
proofs are for \cref{lemma:inv-preserve,lemma:strong-commute,lemma:simatk-dg},
which required machinery for handling our zipping semantics. 
These techniques should be useful and straightforward to build upon in the
future. The Claude AI aided in proving a few minor lemmas and writing tactics,
but all lemma and theorem statements were written by hand.

\iffinalformat
All proofs are available at \url{https://github.com/kssuraaj28/demystify-ifc-rocq} .
\relax\else
The formalization is available at \url{https://anonymous.4open.science/r/demystify-ifc-rocq-8CB2/} .
\fi

\section{Discussion}
\label{sec:Discussion}

We discuss some apparent and real limitations of our approach.

\paragraph{Probability}\label{par:Probability}
The real--ideal paradigm's original formulation is in terms of probabilistic
interactive Turing machines, so dealing with only deterministic adversaries
in \impthreat might seem like a limitation. It is not. It is known that
defending against deterministic adversaries suffices for deterministic
programs~\cite{clarkH08}: a probabilistic adversary is a randomized mixture
of deterministic adversaries, and we consider \emph{all} deterministic
adversaries. This insight is similar to the well-known \emph{dummy-adversary}
theorem~\cite{UC} in cryptography.

\paragraph{Termination sensitivity}\label{par:Termination sensitivity}

The type system only ensures termination-insensitive equivalence. While this is
standard in the literature, it is not what we want for modular security,
because the resulting notion of equivalence is not transitive. The standard fix
is to assume that all (well-typed) programs terminate, which might be relaxed
to only disallow nontermination with untrusted control flow, but we leave this
for future work. Additionally, termination sensitivity subsumes our
fair ideal attacker (\cref{def:fairidlatk}) requirement.

\paragraph{Alternative label models}\label{par:Alternative label models}

Our label model (\cref{sec:heterogeneous NMIFC}) is deliberately
simple.  Richer label models incorporate logical formulas,
computational specifications, and delegation~(e.g., \cite{cm04, cm05,
lz05, flowlocks}). Regardless, \cref{lemma:nmifc-project} is a good
soundness theorem for other label models; any model satisfying that
lemma enjoys our real--ideal result.

\section{Related work}\label{sec:Related work}

The real--ideal paradigm~\cite{goldwasser-micali84, RSIM, constructivecrypto,
Maurer2011AbstractC, UC-IITM, UC-GNUC, SUC, UC-JUC} arose as an alternative to
standard cryptographic game-based security
definitions~\cite{shoup2004-seqgame}. Beyond cryptography, real--ideal style
definitions are increasingly used---mostly implicitly---to give security
specifications for several systems such as hardware~\cite{knox-athalye,
relativesecurity}, operating systems~\cite{UC-OS}, and distributed
systems~\cite{sankbattleship}. However, these works are limited: either
focusing primarily on integrity concerns or having a methodology customized for
their specific use case.

Our work builds on a rich history of language-based security research.
Robust declassification's original formulation~\cite{zm01b} gives it an elegant
semantic definition quite similar to ours. While we require that every active
attack is equivalent to \emph{some} passive attack, the original definition
explicitly constructs a passive attack. Unfortunately, this approach
does not seem to generalize as one introduces transparent endorsement and
nonmalleable information flow~\cite{nmifc}, where this definition becomes
intensional and must exclude `inept' attacks from consideration.
\citet{soloviev2023security} give an \emph{epistemic logic} framework for
defining robust declassification and transparent endorsement. Their treatment
is complicated and not mechanized, and it is unclear whether it generalizes to
nonmalleable information flow. We believe that the real--ideal paradigm is the
better characterization of security, even when the security properties involve
`knowledge', much like the state-of-the-art security specifications for
zero-knowledge proofs (ZKPs)~\cite{Lindell17}. \citet{cecchetti-nmpl}
introduces \emph{nonmalleable progress leakage}, a generalization of NMIF to
include progress sensitivity. This definition is also
intensional, requiring downgrading events in program traces.
Viaduct~\cite{viaduct-formal}, which  we discussed in \cref{sub:Connection to Viaduct},
served as a major inspiration for this paper. \citet{ps03} present
information flow analysis for an ML language. Our proof technique for reducing
bisimulation to invariant preservation is similar to theirs.

\emph{Robust hyperproperty preservation} (RHP)~\cite{abateB0HPT19} requires
secure compilers to preserve hyperproperties between the source and target
programs in all contexts. \citet{rhp-uc} recognize the equivalence between RHP
and the universally composable variant of the real--ideal paradigm~\cite{UC}.
\citet{fullyabstractcomp} present an RHP compiler for a source language without
exceptions to a target language with exceptions. Their proof is not
mechanically verified. Importantly, all existing RHP work do not formalize
complex scenarios where attackers may learn secrets, or heterogeneous security
scenarios with mutually untrusting principals.

\section{Conclusion}\label{sec:Conclusion}

This work presents the real--ideal paradigm as a simple, widely applicable, and
useful framework for specifying and reasoning about language-based security.
While we are not the first to argue for this paradigm, we believe that this
work is the most general and convincing language-based demonstration of the
idea, showing that it works for reasoning about both confidentiality and
integrity, \emph{and} their interactions, for \emph{all} programs written in a
language allowing both secret and untrusted control flow. We show several
applications of this paradigm, from elegant reformulations of existing security
properties to a novel real--ideal soundness theorem for information flow type
systems. Formulating information flow type soundness in the real--ideal
paradigm yields important insights about why it is useful for building
distributed systems with heterogeneous security. We formally verify an NMIFC
type checker offering such guarantees. The generality of the language and the
properties being considered required development of new proof techniques, which
are all fully mechanized in Rocq.

AI is making it cheaper to produce high-quality---even formally
verified---software. But formal verification is only useful when we know
\emph{the right specifications} to verify the software against.  We believe
this work is an important step in this direction.

There are many future directions for extensions to this work. We do
not consider probabilistic programs here, but adding probability as a
first-class primitive to \impthreat would be a good step in this
direction.  The \emph{universal composability}~\cite{UC} variant of
the real--ideal paradigm comes with strong \emph{compositionality
theorems}, enabling designers to build larger systems from simpler
ones. It would be interesting to explore how to connect this notion of
compositionality with existing language-based compositionality
mechanisms.  Our heterogeneous security definition assumes
\emph{static corruption} and \emph{static trust}, in which adversarial
power and trust relationships among principals are fixed at the start
of execution.  
Generalizing our approach to handle \emph{dynamic corruption} and \emph{dynamic
delegation}, where adversarial power and trust evolve as the system runs, is
promising future work.

\ifacknowledgments
\section*{Acknowledgments}

Dave Naumann, Hanxi Chen, Silei Ren, and the anonymous reviewers gave us useful
feedback on the paper.

Any opinions, findings, conclusions, or recommendations expressed
here are those of the author(s) and do not necessarily reflect those
of these sponsors.
\fi

\finalpage

\paperappendix

\section{NMIF soundness proof sketch}
\label{sec:NMIF soundness}

\subsection{The simulator construction}
\label{sub:The simulator construction}

\Cref {thm:simplenmifcsound} is a $\forall \prog.\ \forall \realatk.\
\exists \idealatk. \dots $ statement, which requires that we construct
the simulator \simatk from \prog and \realatk. Recall that an adversary
is defined  as a triple: 
$\langle \stadvset,\ \initadv,\ \stepadv \rangle$. 
We now describe how these
three components are constructed for our simulator.

\paragraph{The state space}
\label{par:Simulator state space}

\begin{figure}[b]
  \centering
  \begin{syntax}
  \groupleft{
    \abstractCategory[\setfnt{State_{adv}}]{\advst}
    \qquad
    \abstractCategory[\setfnt{Store}]{\sigsim}
  }
  \separate
  \category[\setfnt{Prog_{sim}}]{\pisim}
    \alternative{\epsilon}
    \alternative{[\mvarfnt{c^?}] \,::\, \pisim}
    \alternative{[\mvarfnt{c^?}\, \advdevil]\,::\, \pisim}
  \separate
  \category[\setfnt{State_{sim}}]{\mvarfnt{ss}}
    \alternative{\langle \advst,\ \sigsim,\ \pisim \rangle}
\end{syntax}
     \caption{Simulator data for a real adversary with state space \stadvset}
  \label{fig:sim-data}
\end{figure}

\Cref{fig:sim-data} shows the state space of the simulator \simatk. \simatk's
state \emph{wraps} the state space of the real adversary \realatk with two
components, \sigsim and \pisim.

\sigsim is the simulator's version of the program store. While \realatk is able
to \emph{learn} from the program store's public view
using its \cmdpost function, the simulator is subject to the ideal attacker
restriction (\cref{def:dgonlyadv}), and cannot learn from the program store
(except during \kwfnt{declassify} statements). Thus, \simatk maintains \sigsim\
as a simulator-only store copy as part of its state, which it reads as it tries
to emulate the real adversary. As we show in \cref{sub:Combined configurations},
\simatk ensures that \sigsim's public view matches with the real program store's
public view at all times.

\simatk also maintains \pisim, which can be thought of as its own
copy of the currently executing \emph{system continuation} \syscont\
(\cref{fig:impthreatconfig}). Like \syscont, \pisim elements also contain an
optional command along with a bit describing whether the real adversary runs at
that point. However, in general, $\syscont \ne \pisim$. \pisim only contains
entries when \simatk is active. Even when active, \pisim's entries may be
$\epsilon$ if not required by \simatk. The attacker bit \advdevil marks whether the
\emph{real adversary} \realatk is active at that point, which is required for
both computation of \sigsim as well as adversarial information leaking.

Finally, the simulator maintains a copy of the real attacker \realatk's state
\advst, for a \emph{black-box simulation}~\cite{Lindell17} strategy. The goal here is
to ensure that the simulator's \advst matches with the real adversary's \advst
at all times by feeding it \sigsim. Maintaining \advst thus lets \simatk leak
information exactly how the real attacker \realatk would have.

\paragraph{The initial state construction \mvarfnt{adv_{init}}}\label{par:Simulator init}
The \advst component of the initial state triple $\langle \advst, \sigsim,
\pisim \rangle$ is constructed exactly how \realatk would have. \sigsim is
constructed by copying all the initial store's public variables, which \simatk\
is required to do in order to simulate the rest of \realatk's interactions.
Finally, \simatk constructs \pisim from \prog, with all commands marked with
\advdevil, since \realatk is active in all commands of the initial configuration
(\cref{par:The initial configuration}).

\paragraph{The next-step computation  \mvarfnt{adv_{next}}}\label{par:Simulator next}
\simatk uses \pisim as its internal model of what the true system continuation
\syscont is doing, and performs the computations needed for simulation as
required by the commands \cmd in \pisim.

For a non-downgrade \cmd, \simatk simulates the command--attacker
\emph{sandwiched} semantics (\cref{fig:compose-step}) to compute the new value
of its \sigsim. Similarly, \simatk computes the new \pisim by simulating how
\realatk's requested \emph{substeps} (\cref{fig:zip-step}) are zipped. 
We want \simatk to run whenever \realatk does. Zipping is done with a \prog
that \cmd would have stepped using $\cntarr$ to, except for when \cmd has a
\emph{secret guard}; if \cmd is an \kwfnt{if} / \kwfnt{while} with a boolean
test $\mvarfnt {t} (\mvarfnt{x})$ and \mvarfnt{x} is secret, \simatk does not
simulate \prog's branch (which is impossible in general).

When \cmd is a nontrivial downgrade however, \simatk gets to exert
more influence on \sigidl, the ideal program store. But this influence is very
controlled: explicit corruptions of \kwfnt{endorse} sources, and reading of
only \kwfnt{declassify} destinations.

\subsection{Behavioral equivalence}\label{sub:Proving eq}
We now need to show $ \sembr{\prog \mid \realatk} \wsquigeq \sembr{\prog \mid
\simatk}$. Proving this requires \emph{weak bisimulation}. Let us denote the set
of real and ideal configurations (\cref{fig:impthreatconfig}) using \cfgrealset 
and \cfgidealset respectively. Showing weak bisimulation means constructing a
\emph{bisimulation relation} $\mvarfnt{R} \subseteq \cfgrealset \times \cfgidealset$,
and showing that \mvarfnt{R} agrees with both the initial configurations
of \cfgrealset and \cfgidealset, and is preserved by stepping, which we denote by
\stepreal and \stepideal. For the former, we simply have to show that the
initial configurations are \emph{related} by \mvarfnt{R}.

For the latter we show that, if \configreal and \configideal are related, denoted
$(\configreal)\; \mvarfnt{R}\; (\configideal)$, then only one of three things can
happen:

\begin{enumerate}
    \item 
    $
        (\stepreal\; \configreal) 
        \; 
        \mvarfnt{R}
        \; 
        (\stepideal\; \configideal)
    $ 
    \item 
    $
        (\stepreal\; \configreal) 
        \; 
        \mvarfnt{R}
        \; 
        (\configideal)
    $ 
    \item 
    $
        (\configreal)
        \; 
        \mvarfnt{R}
        \; 
        (\stepideal\; \configideal)
    $ 
\end{enumerate}

That is, while one configuration steps, the other may \emph{stutter}.
Additionally, we must ensure that the generated events (\cref{par:Defining
behavior}) match when both configurations step, and that a stutter step on one
side matched by no step on the other emits an $\epsilon$ event.

$\cfgrealset \times \cfgidealset$ is a complex object, and consists of the real
program, the real adversary, the ideal program, and the ideal adversary
executing concurrently. In order to tame this complexity, we construct a
\emph{combined configuration} from which all necessary components are extracted.
In essence, we reduce the binary problem of bisimulation to a unary problem of
\emph{invariant preservation}.

\subsection{Combined configurations}
\label{sub:Combined configurations}

\begin{figure}
  \centering
  \begin{syntax}
  \categoryFromSet[\pcset]{\pc}{\{\PT, \PU, \ST, \SU\}}
  \separate
  \category[\cmdcmbset]{\cmdcmb}
    \alternative{\langle \cropt,\ \ciopt,\ \pc \rangle}
    \alternative{\langle \cropt,\ \ciopt,\ \pc \rangle\, \advdevil}
  \separate
  \category[\contcmbset]{\kcmb}
    \alternative{\epsilon}
    \alternative{\cmdcmb \,::\, \kcmb}
  \separate
  \category[\stcmbset]{\stcmb}
    \alternative{\langle \sigr,\ \sigi,\ \sigs,\ \advst \rangle}
  \separate
  \category[\cfgcmbset]{\configcmb}
    \alternative{\langle \stcmb,\ \kcmb \rangle}
\end{syntax}
   \caption{Combined configuration}
  \label{fig:cfg-cmb-data}
\end{figure}
An elegant way to construct the bisimulation relation \mvarfnt{R} between
\cfgrealset and \cfgidealset is to think of \configreal and \configideal as being
\emph{extracted} from a single \emph{combined object} \configcmb. In our case,
this pays off because, although \configreal and \configideal could be fairly
different in general, they share many similarities for all \emph{reachable configurations}.
This is for two main reasons. First, both \configreal
and \configideal are constructed from the same program \prog. Second,
\simatk\ itself is functionally derived from \prog and \realatk, and shares a
lot of state with them. The structure of a combined configuration \configcmb is
shown in \cref{fig:cfg-cmb-data}. At a high level, it is similar to a single
configuration (\cref{fig:impthreatconfig}), and contains a \emph{combined state}
\stcmb and a \emph{combined command list} \kcmb. A combined state \stcmb is a
4-tuple containing the \emph{real system store} (\sigr), the \emph{ideal system
store} (\sigi), the \emph{simulator private store} (\sigs), and the \emph{real
adversary state} (\advst).

On the other hand, \kcmb is a list of \emph{combined commands} \cmdcmb.
A combined command may be marked with a \advdevil, which shows that it is
attacked. This single bit suffices to derive both whether \realatk\
is active as well as \simatk is active. Additionally, a \cmdcmb also contains
the currently running real (\cropt) and ideal (\ciopt) commands. The final piece
is \pc, a \emph{ghost variable}~\cite{verifyingparallel,ab04}, which we use to require that
\cropt and \ciopt are well-typed (\cref{fig:nmifc-types}).

\paragraph{Combined configuration extraction}\begin{figure}
  \centering
  \begin{tikzcd}[column sep=huge, row sep=small, arrows={labels={font=\footnotesize}}]
 & & \cfgrealset \\
\cfgcmbset \arrow[r, "\extpair"] & \cfgpairset \arrow[ur, bend left=20, "\extreal"', sloped] \arrow[dr, bend right=20, "\extideal", sloped] & \\
 & & \cfgidealset
\end{tikzcd}
   \caption{Configuration extraction}
  \label{fig:cfg-extract}
\end{figure}

\begin{figure}
  \centering
  \begin{tikzcd}[column sep=huge, row sep=large, arrows={labels={font=\footnotesize}}]
\cfgcmbset \arrow[r, "\extpair"] \arrow[d, "\stepcomb"'] & \cfgpairset \arrow[d, "\steppair"] \\
\cfgcmbset \arrow[r, "\extpair"']                        & \cfgpairset
\end{tikzcd}
   \caption{Extraction and step commute}
  \label{fig:step-comm-pair}
\end{figure}

A combined configuration \configcmb gets \emph{extracted} into the real
configuration (\configreal) and the ideal configuration (\configideal),
as shown in \cref{fig:cfg-extract}. We first use \extpair to obtain a
\emph{configuration pair} $\in \cfgpairset$, which then gets split into the real
and ideal configurations using \extreal and \extideal. A configuration in
\cfgpairset is a pair of \configreal and \configideal, except that their \syscontrl
and \syscontidl are padded to have the same length. \cfgpairset is constructed so
that we don't have to deal with the \emph{weakness} of bisimulation until the
very end of our proof.

The extraction is straightforward except for the construction of \simatk. To
extract \simatk, we need to answer two questions: when does \simatk get to run,
and what is its state?

Simulator \simatk runs whenever the real attack \realatk runs (encoded by
marking \cmdcmb with \advdevil), in order to simulate its corruption and
leakage behaviors. Additionally, it runs whenever there is a real command
\cmdr with a public \pc running. This is required as \cmdr can update public
variables, and \simatk must update its personal \sigsim to match them appropriately.

To explain \simatk's state extraction, it suffices to only explain the
construction of its \pisim, as the other components are straightforward. \pisim\
is \simatk's mental model of the \emph{real system continuation} \syscontrl, and
is very similar to it. The main difference is that every real command \cmdr with
a secret \pc is erased during extraction; \simatk does not need to know it and
cannot know it in general.

\paragraph{\cfgcmbset invariant}\label{par:cmbcfg invariant}
Not all configurations in \cfgcmbset are relevant. We describe the subset of
relevance using a predicate, which is then shown to be an \emph{invariant}.
\begin{definition}[\cfgcmbset invariant]
\label{def:cmb-invariant}
Our invariant, \invcomb, requires the following to hold:
\begin{itemize}
  \item \textbf{State invariant:} The \emph{public views} ($\proj {\sigma}
  {\public}$) of \sigr and \sigs\ agree, and the \emph{trusted views} ($\proj
  {\sigma} {\trusted}$) of \sigr and \sigi agree, where a view of a store
  restricts its domain in the obvious way. That is, $ (\proj {\sigr} {\trusted}
  = \proj {\sigi} {\trusted}) \land (\proj {\sigr} {\public} = \proj {\sigs}
  {\public})$.

  \item \textbf{Command invariant:} For every \cropt in \kcmb, $\pc \wts
  \cropt$. As for \ciopt, if \pc is trusted, $\cropt = \ciopt$. For an
  untrusted \pc, \ciopt must still be well-typed in a public context. That is,
  $\PU \wts \ciopt$.
\end{itemize}
\end{definition}

Maintaining this invariant (\cref{def:cmb-invariant}) is the heart of the
$\wsquigeq$ proof. $(\proj {\sigr} {\trusted} = \proj {\sigi} {\trusted})$ is
necessary for system I/O equivalence, while $(\proj {\sigr} {\public} = \proj
{\sigs} {\public}) $ ensures equivalence of adversary leakage. Additionally,
requiring that $\cropt = \ciopt$ in every trusted \pc ensures the right order
of events.

\paragraph{The \stepcomb function}\label{par:The stepcomb function}
A configuration \configcmb in \cfgcmbset steps
using \stepcomb. As in regular configuration stepping
(\cref{fig:impthreat-semantics,fig:compose-step,fig:zip-step}), the \emph{head
command} \cmdcmb of \kcmb is run at each step. \cmdcmb updates \sigr using
its \cropt, and updates \sigi using its \ciopt. Finally, updates to \sigs are
done exactly how the simulator would have. Computing the
new \kcmb is done by zipping together (\cref{fig:zip-step}) three things: the
real continuation, the ideal continuation, and the real attacker substeps.

We now prove that \cref{def:cmb-invariant} is really an invariant when stepping
with \stepcomb.
\begin{lemma}[Stepping invariant preservation]
\label{lemma:inv-preserve}

\stepcomb preserves \invcomb. That is,
\[
\forall \configcmb.\ \invcomb(\configcmb) \implies \invcomb(\stepcomb(\configcmb))
\]
\end{lemma}

While finding the right invariant was a challenge, the proof of the above lemma
is fairly straightforward. From this point on, we only consider configurations
\configcmb that satisfy \invcomb. We redefine $\cfgcmbset \defined \{ \configcmb \in
\cfgcmbset \mid\ \invcomb (\configcmb) \}$.

\subsection{Extraction commutativity}\label{sub:Extraction commutativity}
\begin{figure}
  \centering
  \begin{tikzcd}[column sep=large, row sep=small, arrows={labels={font=\footnotesize}}]
 & \cfgpairset \arrow[dl, "\extreal", sloped] \arrow[dd, "\steppair"] \arrow[dr, "\extideal", sloped] & \\
\cfgrealset \arrow[dd, "\shortstack{$\stepreal\,/\,\idfn$\\ \circnum{1}}"'] & & \cfgidealset \arrow[dd, "\shortstack{$\stepideal\,/\,\idfn$\\ \circnum{2}}"] \\
 & \cfgpairset \arrow[dl, "\extreal", sloped] \arrow[dr, "\extideal", sloped] & \\
\cfgrealset & & \cfgidealset
\end{tikzcd}
   \caption{Two-step extraction commutativity (\protect\circnum{1} and \protect\circnum{2} cannot both be $\idfn$ simultaneously).}
  \label{fig:step-comm-pair-2}
\end{figure}
\cfgcmbset and \stepcomb are very different from \cfgrealset, \cfgidealset,
\stepreal, and \stepideal. To show that they are still faithful, we prove the
commutativity of \cref{fig:step-comm-pair,fig:step-comm-pair-2}.

\Cref{fig:step-comm-pair} asserts the commutativity of the step and extract
functions that involve \cfgcmbset and \cfgpairset. \steppair describes how
configuration pairs in \cfgpairset step. It ensures that the continuations of
\configreal and \configideal have the same lengths by zipping them together.

\begin{lemma}[Strong extract-step commutativity]
\label{lemma:strong-commute}
\Cref{fig:step-comm-pair} commutes.
\end{lemma}

We deal with the weakness of bisimulation in \cref{fig:step-comm-pair-2},
which asserts that whenever a \configpair steps, either both of its extracted
\configreal and \configideal step, or one \emph{stutters} while the other
steps.

\begin{lemma}[Weak extract-step commutativity]
\label{lemma:weak-commute}
One of the three possibilities of \cref{fig:step-comm-pair-2} commutes.
\end{lemma}

These possibilities match up exactly with the possibilities mentioned in
\cref{sub:Proving eq}.

\subsection{The bisimulation relation}\label{sub:The bisimulation relation}
The bisimulation relation $(\configreal \mathrel{\mvarfnt{R}}\ \configideal)$ is defined as:
  \[
  \exists \configcmb.\;
  \begin{cases}
  \configreal  = \extreal(\extpair(\configcmb)) \\
  \configideal = \extideal(\extpair(\configcmb))
  \end{cases}
  \]

Every \configcmb step gives a new existential witness $\configcmb' = \stepcomb
(\configcmb)$. By \cref{lemma:weak-commute,lemma:strong-commute}, we can show
that exactly one of the three weak bisimulation possibilities (\cref{sub:Proving
eq}) holds. Moreover, although we have not explicitly defined \cfgcmbset's
\emph{event behavior}, it is not very surprising, and does not cause much
complication. Similarly, showing that the initial configurations are related by
\mvarfnt{R} is straightforward. This concludes our proof sketch for $\wsquigeq$.

\subsection{The simulator is downgrade-restricted}\label{sub:The simulator is downgrade-restricted}
We now show that \simatk satisfies \cref{def:dgonlyadv}. The simulator
algorithm is very constrained, and only interacts with \prog at downgrading
points. However, the subtlety is ensuring that \simatk's beliefs about
downgrading (in \pisim) match with reality (\syscontidl). 

This requires an invariant. Our proof
reuses the existing machinery from \cref{sub:Proving eq}. We use \configcmb to
construct the following invariant:
\[
  \exists \configcmb.\;
  \configideal = \extideal(\extpair(\configcmb))
\] 

Unfortunately, we cannot use the \stepcomb function to make \configcmb step
in this case. This is because it does not necessarily make \configideal step as
it could cause \configreal to step. Instead, we use $\stepcomb \circ
\filterideal$, where \filterideal filters out all \cmdcmb with a \emph{non-running}
\ciopt from \configcmb. 

\begin{lemma}[Simulator downgrade-restriction]
\label{lemma:simatk-dg}
\simatk only interacts with \prog at downgrade points. That is, $\simatk \in \dgatkset$.
\end{lemma}

\subsection{Simulator fairness}\label{sub:Simulator fairness}
Fairness (\cref{def:fairidlatk}) requires that \simatk take no unnecessary
steps. This is ensured by our combined configuration extraction function
\extpair. \simatk is active only in two cases: when \realatk is active, or
when \cropt is present with a public \pc. Hence no reachable
configuration has \simatk active while nothing else runs.

\begin{lemma}[Simulator fairness]
\label{lemma:simatk-fair}
\simatk is fair (\cref{def:fairidlatk}) for \prog and \realatk.
\end{lemma}
\checkpage{appendix}{18}
 \end{document}